\documentclass[showpacs,preprintnumbers,amsmath,amssymb,floatfix]{revtex4}
\usepackage[german,english]{babel}
\usepackage{graphicx}
\usepackage{amsmath}
\usepackage{amsfonts}
\usepackage{amssymb}
\usepackage{epsfig}
\usepackage[hang,nooneline]{subfigure}
\newcommand{\insertplot}[5]{\begin{figure}
 \hfill\hbox to 0.05in{\vbox to #5in{\vfill
 \inputplot{#1}{#4}{#5}}\hfill}
 \hfill\vspace{-.1in}
 \caption{#2}\label{#3}
 \end{figure}}
\newcommand{\inputplot}[3]{
 \special{ps: plotfile #1}
}
 
\newcounter{fig}

\begin{document}

\title{\bf Boson Stars with Nontrivial Topology}

\author{{\bf Vladimir Dzhunushaliev$^{1,2,3,4}$}}
\email[{\it Email:}]{v.dzhunushaliev@gmail.com}
\author{{\bf Vladimir Folomeev$^{1,3}$}}
\email[{\it Email:}]{vfolomeev@mail.ru}
\author{{\bf Christian Hoffmann $^1$}}
\email[{\it Email:}]{christian.hoffmann@uni-oldenburg.de}
\author{{\bf Burkhard Kleihaus $^1$}}
\email[{\it Email:}]{b.kleihaus@uni-oldenburg.de}
\author{{\bf Jutta Kunz$1^1$}}
\email[{\it Email:}]{jutta.kunz@uni-oldenburg.de}
\affiliation{$^1$
Institut f\"ur Physik, Universit\"at Oldenburg, Postfach 2503,
D-26111 Oldenburg, Germany\\
$^2$
Dept. Theor. and Nucl. Phys., KazNU, Almaty, 010008, Kazakhstan\\
$^3$Institute of Physicotechnical Problems and Material Science of the NAS
of the
Kyrgyz Republic, 265 a, Chui Street, Bishkek, 720071,  Kyrgyz Republic\\
$^4$ Institute of Experimental and Theoretical Physics,
Al-Farabi Kazakh National University, Almaty 050040, Kazakhstan
}

\date{\today}
\pacs{04.20.JB, 04.40.-b}

\begin{abstract}
We construct boson star solutions in the presence of a phantom field,
allowing for a nontrivial topology of the solutions.
The wormholes residing at the core of the configurations
lead to a number of qualitative changes of the boson star solutions.
In particular, the typical spiraling dependence of
the mass and the particle number on the frequency of the boson stars is lost.
Instead, the boson stars with nontrivial topology
approach a singular configuration in the limit of vanishing frequency.
Depending on the value of the coupling constant,
the wormhole geometry changes from a single throat configuration
to a double throat configuration, featuring a belly inbetween 
the two throats.
Depending on the mass of the boson field and its self-interaction,
the mass and the size of these objects cover many orders of magnitude,
making them amenable to various astrophysical observations.
A stability analysis reveals, that the unstable
mode of the Ellis wormhole is retained in the presence of the
bosonic matter. However, the negative eigenvalue
can get very close to zero,
by tuning the parameters of the self-interaction potential appropriately.
\end{abstract}

\maketitle

\section{Introduction}

Astrophysical compact objects of high observational
and theoretical interest comprise white dwarfs,
neutron stars and black holes \cite{Shapiro:1983du}.
Besides these well-established objects,
however, also more speculative astrophysical objects
like wormholes \cite{Visser:1995cc} 
are met with increasing interest,
both from a theoretical and observational point of view.

Recently, for instance, wormholes have been searched for observationally
\cite{Abe:2010ap,Toki:2011zu,Takahashi:2013jqa},
by looking for their predicted signatures as gravitational lenses, 
first considered in \cite{Cramer:1994qj,Perlick:2003vg}.
On the theoretical side, in particular, their Einstein rings
\cite{Tsukamoto:2012xs}
and their shadows
\cite{Bambi:2013nla,Nedkova:2013msa}
have been studied.
But also mixed systems,
consisting of neutron stars harbouring wormholes at their core,
and their possible astrophysical signatures have been addressed
\cite{Dzhunushaliev:2011xx,Dzhunushaliev:2012ke,Dzhunushaliev:2013lna,Dzhunushaliev:2014mza}.
Recently, also quark matter has been considered
\cite{Harko:2014oua}.

In Einstein gravity the nontrivial topology of wormholes and mixed systems
requires a violation of the energy conditions, that can be accomplished
by the presence of a phantom field, as first employed by Ellis
\cite{Ellis:1973yv,Ellis:1979bh,Bronnikov:1973fh,Morris:1988cz,Morris:1988tu,Lobo:2005us}. 
Since {\sl dark energy} represents the major component of the
Universe today, cosmology suggests that a phantom field might indeed exist.

Besides the dark energy the Universe contains a large amount of 
{\sl dark matter}. Theoretical candidates for dark matter are ubiquitous,
and many suggestions involve new scalar fields.
Some such scalar fields could form extended bound objects,
with sizes varying over many orders of magnitude, 
ranging from microscopic particles to huge galactic halos.
Dubbed boson stars
\cite{Lee:1991ax,Jetzer:1991jr,Liddle:1993ha,Mielke:1997re,Mielke:2000mh,Schunck:2003kk,Broderick:2005xa,Schunck:2008xz},
such extended compact objects might even mimick black holes.

Here we consider mixed systems, consisting of boson stars
with wormholes at their core and investigate their properties.
On the one hand, these systems are simpler than the fermionic
mixed systems studied before.
On the other hand, they still allow for a lot of freedom, since the
properties of the boson fields are not known yet.
In particular, the mass of the boson fields is a free parameter,
and so are the form and the strength of their self-interaction.

Following earlier studies
\cite{Mielke:1980sa,Lee:1991ax,Kleihaus:2005me,Kleihaus:2007vk,Kleihaus:2011sx}
we here choose a rather general self-interaction potential for the
boson field, which possesses besides a mass term, also a quartic
and a sextic term.
A further advantage of such a self-interaction is the existence of
localized finite energy solutions, when gravity is turned off.
Indeed, in the probe limit the solutions form non-topological solitons
or $Q$-balls \cite{Friedberg:1976me,Coleman:1985ki,Friedberg:1986tq,Volkov:2002aj}.
These exist only in a finite frequency range of the boson field.

$Q$-balls may also be considered as solutions of the field
equations obtained in the probe limit, where the backreaction
of the boson field on the metric is neglected. 
Clearly, the simplest background metric to obtain $Q$-balls
is the Minkowski space-time.
Here we first consider boson star solutions with a nontrivial
topology in the probe limit. We therefore take the Ellis wormhole
as the background metric, to solve for the boson field.
Interestingly, the nontrivial topology does not change the
frequency range, where the solutions exist.

We then take the backreaction into account 
and solve the full set of coupled equations
for the metric, the boson field and the phantom field,
and consider the coupling constant as a parameter.
In the case of a trivial topology, the families of boson star
solutions exhibit characteristic spirals, when 
the dependence of the mass and the particle number on the 
frequency of the boson field is considered.
We here show, that the frequency dependence of the solutions
is drastically changed due to their nontrivial topology.
In particular, the spirals unwind, an effect observed previously
for boson stars in Einstein-Gauss-Bonnet theory
\cite{Hartmann:2013tca}.
Another interesting effect arising from the backreaction is
the occurrence of solutions with a double throat.

By varying the parameters we map the domain of existence
of these families of solutions and chart their physical properties.
We address the emergence and the properties of limiting solutions,
and consider some of their astrophysically relevant characteristics. 
Here a very important question is, of course, the stability
of the solutions.
Therefore we perform a stability analysis, where
we focus on the crucial mode for solutions
with nontrivial topology, 
which is associated with radial perturbations.

In contrast to early work 
\cite{Kodama:1978dw,ArmendarizPicon:2002km},
which seemed to indicate that 
isolated phantom field wormholes could be stable,
later work revealed their instability \cite{Shinkai:2002gv,Gonzalez:2008xk},
in fact these solutions possess an unstable radial mode
\cite{Gonzalez:2008wd}.
Recently it was shown that
this instability of isolated phantom field wormholes is inherited 
by a number of static solutions with nontrivial topology which involve further
matter fields and also by static neutron stars which harbour
wormholes at their core
\cite{Bronnikov:2011if,Dzhunushaliev:2013lna,Charalampidis:2013ixa}.
Here we show, that the instability is retained also by the families
of stationary boson stars with nontrivial topology. 
However, we find that the eigenvalue of the unstable mode can
get very close to zero.

We present in section II
the action, the Ans\"atze and the field equations.
We discuss the numerical results in section III,
investigate the stability with respect to radial perturbations 
in section IV, and give our conclusions in section V.

\section{Action and Field Equations}

\subsection{Action}

We consider Einstein gravity coupled to  a complex scalar field $\Phi$
and a phantom field $\Psi$.
The action 
\begin{equation}
S=\int \left[ \frac{1}{4\pi\alpha}{\cal R} + 
{\cal L}_{\rm ph} +{\cal  L}_{\rm bs} \right] \sqrt{-g}\  d^4x  
 \label{action}
\end{equation}
then consists of the Einstein-Hilbert action
with curvature scalar $\cal R$, coupling constant $\alpha$ 
and determinant of the metric $g$, together with the matter contributions,
the Lagrangian ${\cal L}_{\rm ph}$ of the phantom field $\Psi$,
\begin{equation}
 {\cal L}_{\rm ph} = \frac{1}{2}\partial_\mu \Psi\partial^\mu \Psi \ ,
\label{lpsi}
\end{equation}
and the Lagrangian ${\cal L}_{\rm bs}$ of the complex scalar field $\Phi$
\begin{equation}
{\cal L}_{\rm bs} = 
-\frac{1}{2} g^{\mu\nu}\left( \Phi_{, \, \mu}^* \Phi_{, \, \nu} + \Phi _
{, \, \nu}^* \Phi _{, \, \mu} \right) - U( \left| \Phi \right|) \ ,
\label{lphi}
\end{equation} 
where the asterisk denotes complex conjugation,
\begin{equation}
\Phi_{,{\mu}}  = \frac{\partial \Phi}{ \partial x^{\mu}}
 \ ,
\end{equation}
while $U$ denotes the potential
\begin{equation}
U(|\Phi|) =  \lambda |\Phi|^2 \left( |\Phi|^4 -a |\Phi|^2 +b \right)
\ . \label{U} 
\end{equation}
The potential has a minimum at $\Phi =0$, where $U(0)=0$, 
and a second minimum at some finite value of $|\Phi|$.
The mass of the bosons is given by $m_b=\sqrt{\lambda b}$.
In a Minkowski background
the potential allows for nontopological soliton solutions
\cite{Friedberg:1976me,Friedberg:1986tq}, also referred to as $Q$-balls
\cite{Coleman:1985ki}.

Variation of the action with respect to the metric
leads to the Einstein equations
\begin{equation}
G_{\mu\nu}= {\cal R}_{\mu\nu}-\frac{1}{2}g_{\mu\nu}{\cal R} = 2 \alpha T_{\mu\nu}
\label{ee} 
\end{equation}
with stress-energy tensor
\begin{equation}
T_{\mu\nu} = g_{\mu\nu}{{\cal L}}_M
-2 \frac{\partial {{\cal L}}_M}{\partial g^{\mu\nu}} \ ,
\label{tmunu} 
\end{equation}
where ${\cal L}_{\rm M} = 
{\cal L}_{\rm ph}+{\cal L}_{\rm bs} $ is the matter Lagrangian.

\subsection{Ans\"atze}

An appropriate choice for the line element 
of spherically symmetric solutions
with a wormhole at their core is given by
\begin{equation}
ds^2 = -A^2 dt^2 +d\eta^2 + R^2d\Omega^2 \ ,
\label{lineel}
\end{equation}
where $d\Omega^2=d\theta^2 +\sin^2\theta d\varphi^2$ 
denotes the metric of the unit sphere, while $A$ and $R$ are functions of 
$\eta$.
The radial coordinate $\eta$ takes positive and negative 
values, i.e. $-\infty< \eta < \infty$. 
The two limits $\eta\to \pm\infty$ then
correspond to two distinct asymptotically flat regions.

We parametrize the complex scalar field $\Phi$ via
\begin{equation}
\Phi 
  =  \phi (r) ~ e^{ i\omega t } \ ,   \label{ansatzp}
\end{equation}
where $\phi (r)$ is a real function,
and $\omega$ denotes a frequency.
The phantom field $\Psi$ has only a dependence on the radial coordinate,
\begin{equation}
\Psi 
=  \psi (r) \ .
\label{ansatzph}
\end{equation}

\subsection{Einstein and Matter Field Equations}

Substitution of the above Ans\"atze into the Einstein equations 
$G_\mu^\nu=2\alpha T_\mu^\nu$ yields
\begin{eqnarray}
\frac{2 R R'' +R'^2 -1 }{R^2}
 & =  &  
\alpha\left[-2 \phi'^2-2 \omega^2\frac{\phi^2}{A^2}-2 U(\phi)+\psi'^2\right] \ ,
\label{eeq_tt}\\
\frac{A R'^2 -A + 2R R' A'}{AR^2}
 & = & 
\alpha\left[2 \phi'^2+2 \omega^2\frac{\phi^2}{A^2}-2 U(\phi)-\psi'^2\right] \ ,
\label{eeq_rr}\\
\frac{A R'' + A' R' +A'' R}{AR} 
 & = &
\alpha \left[-2 \phi'^2+2 \omega^2\frac{\phi^2}{A^2}-2 U(\phi)+\psi'^2\right]
\label{eeq_oo}			
\end{eqnarray}
for the $tt$, $\eta\eta$ and $\theta\theta$ components, respectively.

The equations for the functions of the 
complex scalar field and the phantom field 
are obtained from the variation of the action with respect to $\phi$
and $\psi$, respectively.
They read 
\begin{eqnarray}
\left[A R^2\phi' \right]'
& = & -\omega^2\frac{R^2}{A}\phi+AR^2 \frac{1}{2}\frac{dU}{d\phi} \ ,
\label{eqSk}\\
\left[ A R^2 \psi'\right]' &=& 0\ .
\label{eqPh}
\end{eqnarray}
Integrating the last equation we obtain
\begin{equation}
\psi' = \frac{D}{A R^2} \ ,
\label{phip}
\end{equation}
where the constant $D$ represents the scalar charge of the phantom field.
By substituting $\psi'^2 = D^2/A^2 R^4$ in the Einstein equations,
the phantom field can be eliminated from the 
remaining set of equations.

%
%

By adding Eq.~(\ref{eeq_rr}) to Eq.~(\ref{eeq_tt}) 
and to Eq.~(\ref{eeq_oo}), we eliminate the $\phi'^2$ term 
and the $\psi'^2$ term.
The final set of Einstein equations to be solved can then be cast in the form
\begin{eqnarray}
R'' & = & \frac{A - A R'^2 -R R'A'}{AR} 
- 2\alpha R U  \ ,
\label{eqR}\\
A'' & = & -\frac{2 R'A'}{R} 
 - 2\alpha A \left[U -2\omega^2\frac{\phi^2}{A^2}\right]  \ .
\label{eqA}
\end{eqnarray}
Together with Eq.~(\ref{eqSk}) they form a system of second order
ODEs to be solved numerically.

\subsection{Global charges}

The mass $M$ of the stationary asymptotically flat solutions
can be obtained from the Komar expression \cite{Wald:1984},
\begin{equation}
{M} = 
 \frac{1}{{4\pi}} \int_{\Sigma}
 R_{\mu\nu}n^\mu\xi^\nu dV
\ . \label{komarM1}
\end{equation}
Here $\Sigma$ denotes an asymptotically flat spacelike hypersurface,
$n^\mu$ is normal to $\Sigma$ with $n_\mu n^\mu = -1$,
$dV$ is the natural volume element on $\Sigma$,
and $\xi$ denotes an asymptotically timelike Killing vector field
\cite{Wald:1984}.
The mass $M$ can be read off
directly from the asymptotic expansion of the 
metric component $g_{tt}$
\begin{equation}
g_{tt} \longrightarrow -1+\frac{2M}{R} 
\ \ \ \Longleftrightarrow  \ \ \
A  \longrightarrow 1-\frac{M}{R} \ .
\end{equation}

The Lagrange density is invariant under the global phase transformation
\begin{equation}
\displaystyle
\Phi \rightarrow \Phi e^{i\chi} \ ,
\end{equation}
leading to the conserved current
\begin{eqnarray}
j^{\mu} & = &  - i \left( \Phi^* \partial^{\mu} \Phi 
 - \Phi \partial^{\mu}\Phi ^* \right) \ , \ \ \
j^{\mu} _{\ ; \, \mu}  =  0 \ .
\end{eqnarray}
The associated conserved charge $Q$ is then obtained from the time-component
of the current,
\begin{eqnarray}
Q &=- & \int j^t \left| g \right|^{1/2} d\eta d\theta d\varphi 
\nonumber \\
&=& 8 \pi \omega \int_0^{\infty} |g| ^{1/2}   \frac{\phi^2}{A^2} \,d\eta \, \ . 
\label{Qc}
\end{eqnarray}
The charge $Q$ corresponds to the particle number
of the self-interacting bosons.

The scalar charge $D$ of the phantom field 
can be obtained from Eq.~(\ref{eeq_rr}),
\begin{eqnarray}
\alpha D^2  & = & A^2 R^2\left[1-R'^2-2\frac{A' R R'}{A}\right]
 + 2 \alpha A^2 R^4\left[\phi'^2-U+\omega^2\frac{\phi^2}{A^2}\right] \ .
\label{eqD2}
\end{eqnarray}
We employ the condition $D=const$ to monitor the quality of the
numerical solutions.

\subsection{Throats}

In the following we restrict to symmetric solutions,
i.e., solutions whose metric functions are symmetric under
$\eta \to - \eta$, and whose matter field function are
either symmetric or antisymmetric.

The metric function $R$ may be considered as a circumferential
radial coordinate.
Since we want to obtain solutions with wormholes at their core,
we assume that the function $R$ does not possess a zero.
In asymptotically flat solutions, the function $R$ 
then tends towards $|\eta|$ in the asymptotic regions. 
Consequently, the function $R$ must possess at least one minimum. 

Because of the assumed symmetry, $\eta=0$ must correspond
to an extremum of the function $R$,
and thus $R'(0)=0$.
In the simplest case, the function $R$ will have a single minimum 
located at $\eta=0$, $R(0)=r_0$.
However, the function $R$ could also possess a maximum
at $\eta=0$ and two minima, located symmetrically on each side.
In principle, the function $R$ could possess even more extrema.

To understand the nature of the extremum at $\eta=0$,
we consider Eq.~(\ref{eqR}) at $\eta=0$
\begin{equation}
R''(0) = \frac{1}{r_0} \left[1-2\alpha r_0^2 U\left(\phi(0)\right)\right]
=\frac{1}{r_0} \left(1- \alpha/\alpha_{\rm cr}\right) 
\ ,
\end{equation}
where 
\begin{equation}
\alpha_{\rm cr} =1/(2 r_0^2 U\left(\phi(0)\right)) \ .
\label{acrit}
\end{equation}
Thus, $R(0)$ is a minimum when $\alpha < \alpha_{\rm cr}$,
whereas it is a maximum when $\alpha > \alpha_{\rm cr}$.
Here the first case $\alpha < \alpha_{\rm cr}$ represents the simplest
wormhole scenario, where a surface of minimal area separates two 
asymptotically flat regions. 
In the second case, however, $R(0)$ is a local maximum,
and thus represents a surface of maximal area, i.e., an equator.
This then implies that there are (at least) two minima of $R$, 
one for $\eta < 0$ and a second symmetric one for $\eta > 0$.
In the case of two such minima, the wormhole possesses a double throat,
with an equator located symmetrically inbetween at $\eta=0$.

The area of a throat ${\cal A}_{\rm th}$ is determined
by
\begin{equation}
{\cal A}_{\rm th} = 4 \pi R^2(\eta_{\rm th}) \ .
\end{equation}
When a wormhole has a single throat, its location is at
$\eta_{\rm th}=0$, and its area is given by
\begin{equation}
{\cal A}_{\rm th} = 4 \pi r_0^2  \ .
\end{equation}
When a wormhole has a double throat, on the other hand, 
$\eta=0$ corresponds to an equator,
and the locations of the two throats are at 
finite values, $\pm \eta_{\rm th}$.

The surface gravity $\kappa$ of the throat can be evaluated via
\begin{equation}
\kappa^2 = -1/2 \left. 
(\nabla_\mu \xi_\nu)(\nabla^\mu \xi^\nu) \right|_{\eta_{\rm th}} \ 
\end{equation}
with the timelike Killing vector field $\xi^\mu$.
For the spherically symmetric metric employed,
the surface gravity is then given by
\begin{equation}
\kappa =  A'(\eta_{\rm th}) \ .
\label{kap}
\end{equation}
Consequently, the wormholes with a single throat have vanishing $\kappa$,
while the wormholes with a double throat
possess a finite surface gravity.

\subsection{Boundary Conditions}

At the extremal surface $\eta=0$ - the throat or the equator -
we impose the boundary conditions
\begin{equation}
R(0)= r_0 \ , \ \ \ 
R'(0)= 0 \ , \ \ \ 
A'(0)=0 \ .
\label{bcthro}
\end{equation}
With the first condition we fix the circumferential
radius of the throat or the equator,
while the second condition is simply the extremum condition. 
The third condition is imposed by the symmetry of the solutions
with respect to an interchange of the universes.

In the asymptotic regions
we impose the following boundary conditions
\begin{equation}
A(\eta\to +\infty) \to 1 \  ,  \ \ \ 
\phi(\eta\to -\infty) \to 0 \ ,  \ \ \ 
\phi(\eta\to +\infty) \to 0 \ .
\label{bcasym}
\end{equation}
While the first condition sets the time scale, the 
second condition and  the third condition follow from 
the requirement of finite energy.

\subsection{Energy conditions}

The violation of the null energy condition (NEC)
implies the violation of the weak and the strong energy condition.
Therefore, we address only the NEC, which requires
\begin{equation}
\Xi = T_{\mu\nu} k^\mu k^\nu \ge 0 \ ,
\end{equation}
for all (future-pointing) null vector fields $k^\mu$.

We reexpress this condition by
making use of the Einstein equations,
and then obtain for spherically symmetric solutions 
the new conditions
\begin{equation}
-G_t^t+G_\eta^\eta \geq  0 \ , \ \ \ {\rm and } 
\ \ \ -G_t^t+G_\theta^\theta\geq  0 \ .
\label{Nulleng}
\end{equation}
The null energy condition is violated,
when one or both of these conditions
do not hold in some region of the spacetime considered.
This is the case for all of the solution studied.

\subsection{Units}

Since we would like to consider our results also from an
astrophysical point of view, we need to introduce appropriate
factors of $\hbar$, $G$ and $c$ into the action, and connect
the dimensionful quantities with the corresponding dimensionless quantities,
employed in the calculations.
In dimensionful quantities
the total Lagangian ${\cal L}_{tot}$ then reads
\begin{equation}
{\cal L}_{tot} =\frac{c^4}{16\pi G}{\cal R}
-\frac{\hbar c}{2} 
g^{\mu\nu}\left( \Phi_{, \, \mu}^* \Phi_{, \, \nu} + \Phi _
{, \, \nu}^* \Phi _{, \, \mu} \right) 
- \frac{1}{\hbar c} U( \left| \Phi \right|) 
+\frac{\hbar c}{2}\partial_\mu \Psi\partial^\mu \Psi \ 
 \label{actiondens}
\end{equation}

Now we introduce the associated dimensionless quantities 
via $\Phi = \Phi_0 \hat{\Phi}$,
$\Psi = \Psi_0 \hat{\Psi}$ and $\eta = \eta_0 \hat{\eta}$,
where $\Phi_0$ and $\Psi_0$ have the dimension of an inverse length,
and $\eta_0$ has the dimension of length.
For convenience we rewrite the potential as
\begin{equation}
U(\left| \Phi \right|) = (m_0 c^2)^2 \Phi_0^2 
\left[| \hat{\Phi} |^2 
         + c_4 | \hat{\Phi} |^4+ c_6 | \hat{\Phi} |^6\right]
=  (m_0 c^2)^2 \Phi_0^2\, \hat{U}	(| \hat{\Phi}|) 
\end{equation}
The Lagrangian then assumes the form
\begin{equation}
{\cal L}_{tot} =\frac{c^4}{8\pi G\eta_0^2}\left\{ \frac{\hat{\cal R}}{2}
-2\alpha \left[  \frac{1}{2} g^{\mu\nu}
  \left( \hat{\Phi}_{, \, \mu}^* \hat{\Phi}_{, \, \nu} 
       + \hat{\Phi}_{, \, \nu}^* \hat{\Phi}_{, \, \mu} \right) 
       +\left(\frac{m_0 c^2}{\hbar c}\right)^2 \eta_0^2 \, \hat{U}( |\hat{\Phi} |)
-\frac{1}{2}\partial_\mu \hat{\Psi}\partial^\mu {\hat \Psi} \right]\right\} \ ,
 \label{actiondenssca}
\end{equation}
with the coupling constant $\alpha$ proportional to Newtons's constant,
\begin{equation}
\alpha = \frac{4\pi G}{c^4} \hbar c \Phi_0^2,
\end{equation} 
and we made the choice $\Psi_0=\Phi_0$.

Next we fix the length scale $\eta_0$ by 
$\eta_0 = G M_0/c^2$, such that the mass $M$ is given in units of $M_0$.
In order to compare with the potential in Eq.~(\ref{U}) we set
\begin{equation}
\left(\frac{m_0 c^2}{\hbar c}\right)^2 \eta_0^2 
\left[| \hat{\Phi} |^2 + c_4 | \hat{\Phi} |^4 + c_6 | \hat{\Phi} |^6\right]
= \lambda b | \hat{\Phi} |^2 - \lambda a | \hat{\Phi} |^4 +\lambda \hat{\Phi} |^6
\ .
 \label{compareU}
\end{equation}
Using 
\begin{equation}
\eta_0 = \hbar c \frac{M_0 c^2}{(m_{\rm Pl} c^2)^2} \ , \ \ \ 
\Phi_0^2 = \frac{\alpha}{4\pi} \left(\frac{m_{\rm Pl} c^2}{\hbar c}\right)^2 \ ,
\end{equation}
with the Planck mass
$m_{\rm Pl} = \sqrt{\hbar c/G}$, we find for the mass of the complex
boson field
\begin{equation}
m_0 = \frac{m_{\rm Pl}^2}{M_0} \sqrt{\lambda b} \ ,
\label{Pmass}
\end{equation}
and for the self-interaction potential
\begin{eqnarray}
U(\left| \Phi \right|)  & = & 
(m_0 c^2)^2 \left| \Phi \right|^2
- a\frac{4 \pi(\hbar c)^2}{\alpha} \left(\frac{m_{\rm Pl}}{M_0}\right)^2
     \left| \Phi \right|^4
+  \left( \frac{4 \pi(\hbar c)^2}{\alpha}\right)^2  \frac{1}{(M_0 c^2)^2}
 \left| \Phi \right|^6
 \nonumber\\
& = & (m_0 c^2)^2 \left| \Phi \right|^2
- a\beta \left(m_{\rm Pl}c^2\right)^2
     \left| \Phi \right|^4
+ \beta^2 \left(M_0 c^2\right)^2
 \left| \Phi \right|^6 \ , 
\end{eqnarray}
with $\beta=\frac{4 \pi}{\alpha}\left(\frac{\hbar c}{M_0 c^2}\right)^2$.
Thus $\alpha$ can be considered to tune the self-interaction.

\section{Numerical Results}

After briefly addressing the numerical method employed,
we first discuss the probe limit of the solutions,
solving for the boson field $\phi$ in the background
of an Ellis wormhole.
For the self-interaction potential $U(\phi)$ we fix
the parameters according to
$\lambda=1$, $a=2$ and $b=1.1$.

Subsequently, we couple gravity and
solve the full system of coupled nonlinear ODEs for the given set of
boundary conditions and the parameters $\alpha$ and $r_0$.
We note, that the quality of the numerical solutions is high.
In particular,
the variation of the constant $D$ as computed from Eq.~(\ref{eqD2})
is typically less than $10^{-9}$.

We here first consider the families of solutions themselves.
In order to fix the scale we choose a value for the throat size $r_0$.
This leaves $\alpha$ as a free parameter as well as the
boson frequency $\omega$.
Varying $\omega$ for a fixed value of $\alpha$, we then obtain a family of
boson star solutions harbouring a wormhole at their core.
Next we consider the dependence of these families of solutions
on the coupling constant $\alpha$.
We demonstrate that all such families
of solutions start from the limiting Ellis wormhole
solution with vanishing boson field $\Phi$
and end in a singular configuration.
Finally, we address some astrophysically relevant
properties of the solutions.

\subsection{Numerical Method}

We employ a collocation method for boundary-value ordinary
differential equations, equipped with an adaptive mesh selection procedure
\cite{COLSYS}.
Typical mesh sizes include $10^3-10^4$ points.
The solutions have a relative accuracy of $10^{-10}$.
The estimates of the relative errors of the global charges
are of order $10^{-6}$.


\begin{figure}[t!]
\begin{center}
\vspace{0.5cm}
\mbox{\hspace{-0.5cm}
\subfigure[][]{\hspace{-1.0cm}
\includegraphics[height=.25\textheight, angle =0]{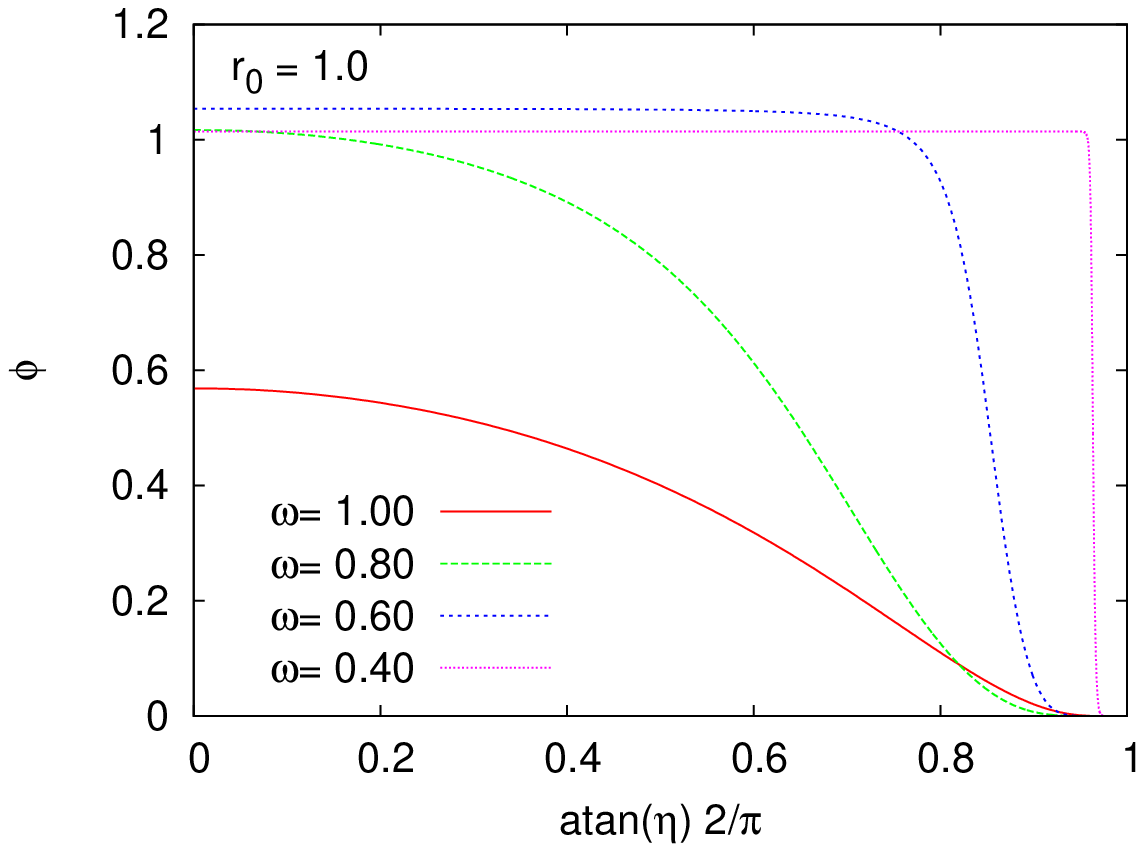}
\label{fig0a}
}
\subfigure[][]{\hspace{-0.5cm}
\includegraphics[height=.25\textheight, angle =0]{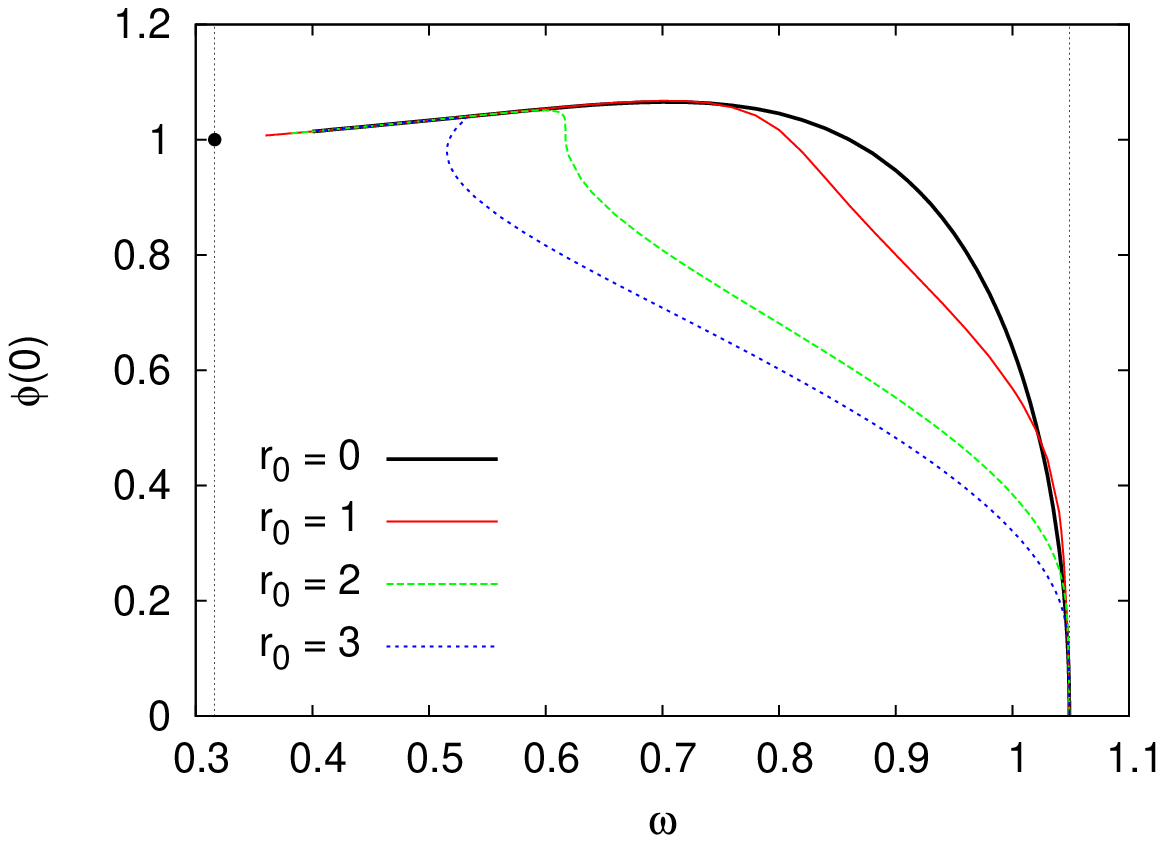}
\label{fig0b}
}
}
\mbox{\hspace{-0.5cm}
\subfigure[][]{\hspace{-1.0cm}
\includegraphics[height=.25\textheight, angle =0]{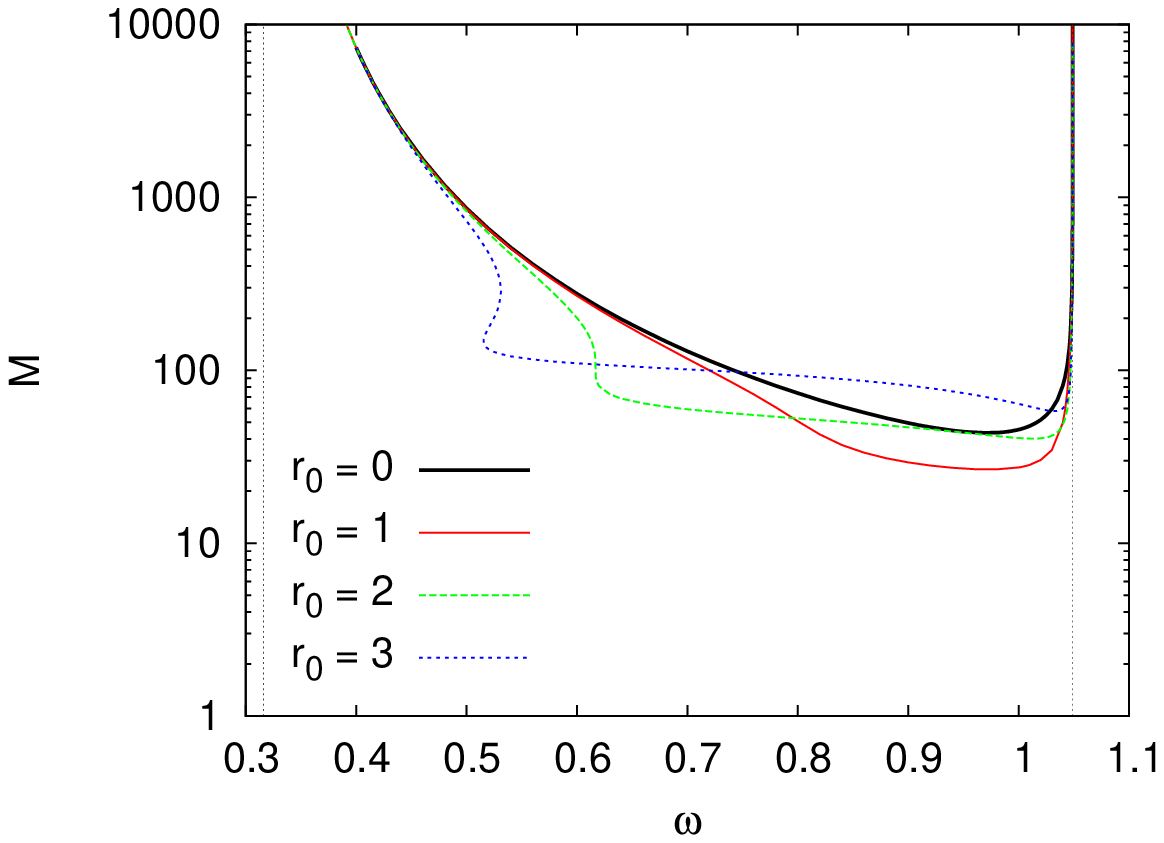}
\label{fig0c}
}
\subfigure[][]{\hspace{-0.5cm}
\includegraphics[height=.25\textheight, angle =0]{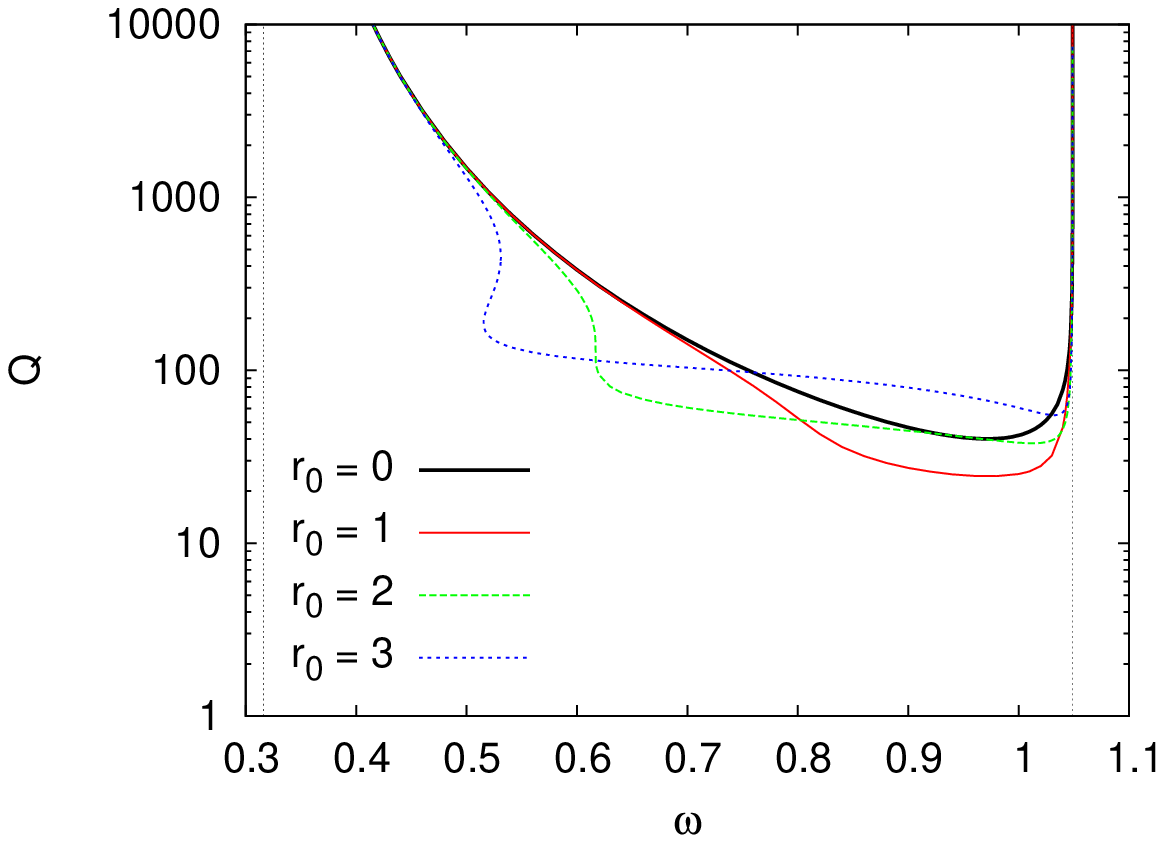}
\label{fig0d}
}
}
\mbox{\hspace{-0.5cm}
\subfigure[][]{\hspace{-1.0cm}
\includegraphics[height=.25\textheight, angle =0]{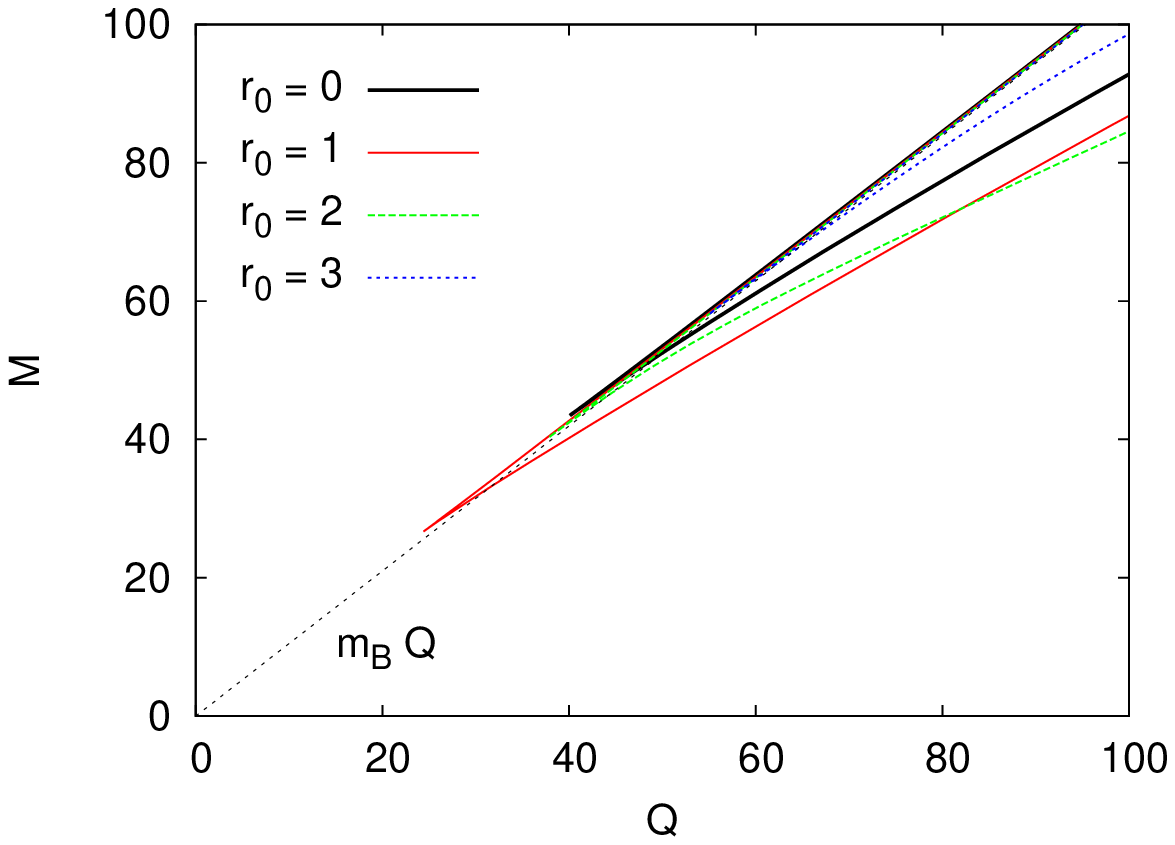}
\label{fig0e}
}
\subfigure[][]{\hspace{-0.5cm}
\includegraphics[height=.25\textheight, angle =0]{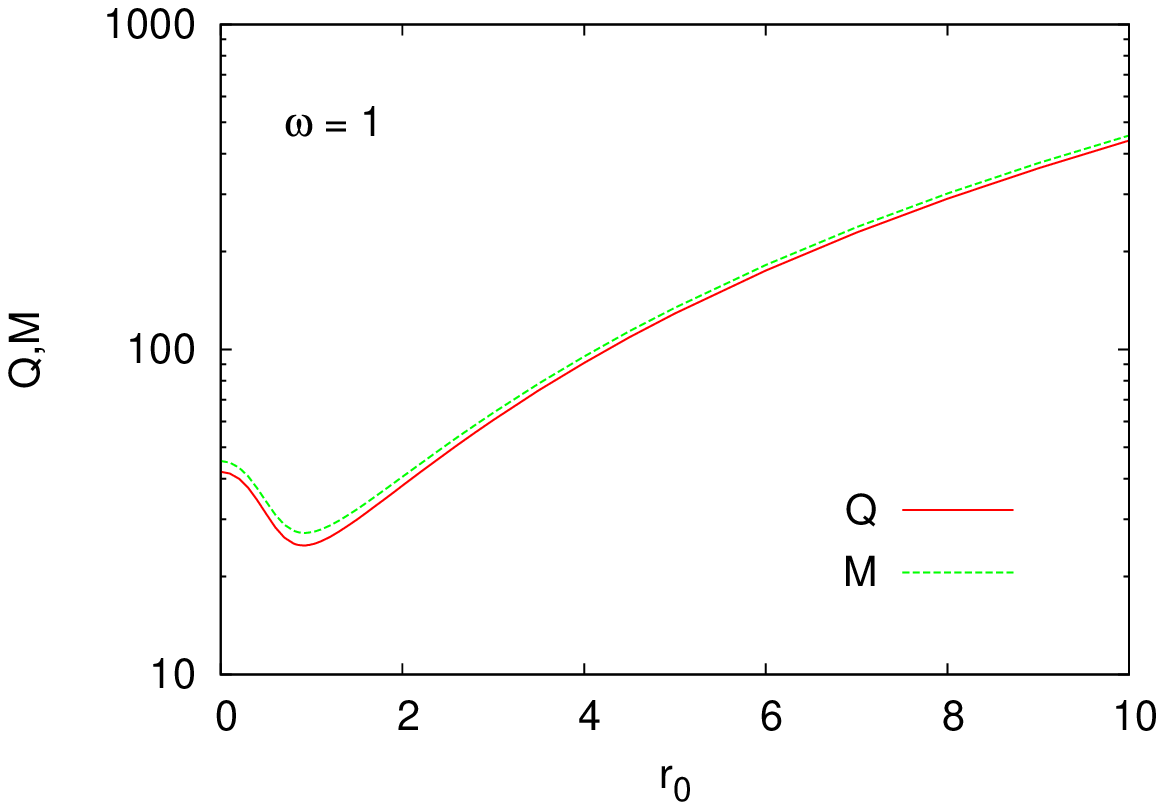}
\label{fig0f}
}
}
\end{center}
\vspace{-0.7cm}
\caption{Probe limit:
(a) The boson field function $\phi$ versus the 
compactified coordinate ${\rm atan}\, \eta$ 
for throat size $r_0=1$ and several values of
the frequency $\omega$.
(b) The value of the boson field function at the throat $\phi_0$
versus the frequency $\omega$ for several
values of the throat size, $r_0=1$, 2, and 3.
The $Q$-ball limit corresponds to $r_0=0$.
The dotted vertical lines represent the minimal and the maximal
values of the frequency $\omega_{\rm min}$ and $\omega_{\rm max}$.
The black dot represents the limiting value of $\phi_0$.
(c) The mass $M$ for the same sets of solutions.
(d) The particle number $Q$ for the same sets of solutions.
(e) The mass $M$ versus the particle number $Q$
for the same sets of solutions. The dotted line
corresponds to the mass of $Q$ free bosons.
(f) The mass $M$ and the particle number $Q$ 
for frequency $\omega=1$
versus the throat size $r_0$.
\label{fig0}
}
\end{figure}

\subsection{Probe Limit}

In the probe limit, the boson field equation is solved in the
background of an Ellis wormhole, where 
\begin{equation}
A=1 \ , \ \ \ R^2=\eta^2+r_0^2 \ .
\end{equation}
Thus a single ODE for 
the function $\phi$
needs to be solved subject to the boundary conditions 
(\ref{bcasym}),
while the frequency $\omega$ and the throat size $r_0$ are varied.
As the throat size tends to zero, the $Q$-ball limit
is reached, where the solutions represent
non-topological soliton solutions in the Minkowski space-time
\cite{Friedberg:1976me,Coleman:1985ki}.

Spherically symmetric $Q$-balls
exist only in a certain frequency range,
$\omega_{\rm min} < \omega < \omega_{\rm max}$
\cite{Friedberg:1976me,Coleman:1985ki,Volkov:2002aj}.
The equation of motion for the scalar field $\phi$
may be viewed as 
effectively describing a particle moving with friction 
in the potential $V(\phi)$,
\begin{eqnarray}
V(\phi) = \frac{1}{2} \, \omega^2 \, \phi^2
 - \frac{1}{2} U(\phi) \ .
\label{Vpot}
\end{eqnarray}
A necessary condition for the existence of $Q$-balls
is then given by $V^{''}(0) < 0$.
This determines the maximal frequency $\omega_{\rm max}$
\begin{equation}
\label{cond1}
\omega^2_{\rm max} =
\frac{1}{2} U''(0) = \lambda \, b = m_b^2 \ .
\end{equation}
Indeed, only for $\omega < \omega_{\rm max}$
the solutions possess an exponential fall-off
at spatial infinity.

The second condition for the existence of $Q$-balls 
is that $V(\phi)$ should
become positive for some nonzero value of $\phi$ 
\cite{Friedberg:1976me,Coleman:1985ki,Volkov:2002aj}.
This yields the minimal frequency $\omega_{\rm min}$,
\begin{equation}
\label{cond2}
\omega^2_{\rm min} =
\min_{\phi} \left[{U(\phi)}/{\phi^2} \right] \; = \; 
\lambda \left(b- \frac{a^2}{4} \right) \ .
\end{equation}

When the Minkowski background is replaced by the Ellis background,
these limits are retained.
Asymptotically, the equation of motion for the boson field 
does not change, and thus the condition $\omega < \omega_{\rm max}$
remains valid.
On the other hand, for $\omega \to  \omega_{\rm min}$,
the boson field function $\phi$ approaches a constant in a large inner region,
i.e., $\phi' \approx 0$,
just as in the case of $Q$-balls.
Thus the deviations of the metric functions from
Minskowski space-time and the presence of the throat become irrelevant.
Consequently, the frequency $\omega$ remains bounded
in the same interval as in Minkowski space-time.

In Fig.~\ref{fig0} we illustrate the solutions 
and their properties in the probe limit.
The boson field function $\phi$ is shown in Fig.~\ref{fig0a} 
for the throat size $r_0=1$ 
and several values of the frequency $\omega$
in the allowed interval $\omega_{\rm min} < \omega < \omega_{\rm max}$.
The behaviour of the boson field function $\phi$ is very similar to
the case of $Q$-balls.
In particular, 
when the frequency $\omega$ approaches its lower limit $\omega_{\rm min}$,
the function $\phi$ tends to a constant in an inner region, 
that increases in size as $\omega \to \omega_{\rm min}$.
At the same time
the value of $\phi$ at the throat, $\phi_0$, tends to a limiting value,
$\phi_0(\omega_{\rm min})=1$,
as indicated in Fig.~\ref{fig0b}. 
(Note, that for $\omega = \omega_{\rm min}$ the field equation is
solved by $\phi(\eta)=1$.)

Fig.~\ref{fig0b} exhibits $\phi_0$ for four sets of solutions,
which correspond to the $Q$-ball case for $r_0=0$, and 
the Ellis background cases with the values for
the throat size $r_0=1$, 
$r_0=2$ and $r_0=3$.
We note that for small frequencies and larger
values of the throat size uniqueness of the solutions is lost.
Here, for a given value of $\omega$, there may be 3 distinct solutions.
Thus instead of a single branch of solutions there arise three branches
of solutions for the larger values of the throat size.

However, for all values of the throat size, the
mass $M$ and the particle number $Q$ of the solutions 
approach those of the $Q$-ball solutions 
when $\omega \to \omega_{\rm max}$
and when $\omega \to \omega_{\rm min}$. 
Thus in these limits, the mass and the particle number
are diverging just as for $Q$-balls.
This is seen in Fig.~\ref{fig0c} and \ref{fig0d},
where the mass $M$ and the particle number $Q$ are exhibited
for the same sets of solutions.
These figures also clearly show the nonuniqueness of the solutions.

The binding energy of the solutions can be extracted from Fig.~\ref{fig0e},
where we show the mass $M$ versus the particle number $Q$
for the same sets of solutions. Here the straight line gives the
mass of $Q$ free bosons,
\begin{equation}
M_{\bf free} = m_b \, Q \ .
\end{equation}
As in the case of $Q$-balls there are always bound solutions,
located on the lower branch, where $M<M_{\bf free}$, and 
unbound solutions, whenever $M>M_{\bf free}$.
For large values of the throat size, the set of bound solutions exhibits
additional structure because of the nonuniqueness.

The dependence of the mass $M$ and the particle number $Q$ on the
throat size in exhibited in Fig.~\ref{fig0f}
for the value of the frequency $\omega=1$.
In the vicinity of this value the mass and the particle number
take their minimal values. 
We observe, that the mass $M$ and the particle number $Q$
do not change monotonically with the throat size, 
as also seen in Fig.~\ref{fig0d}.
To conclude, we observe many similarities between the solutions
in the probe limit in the background of an Ellis wormhole
and the $Q$-ball solutions of Minkowski space-time.

\subsection{Gravitating Solutions}

\begin{figure}[t!]
\begin{center}
\vspace{0.5cm}
\mbox{\hspace{-0.5cm}
\subfigure[][]{\hspace{-1.0cm}
\includegraphics[height=.25\textheight, angle =0]{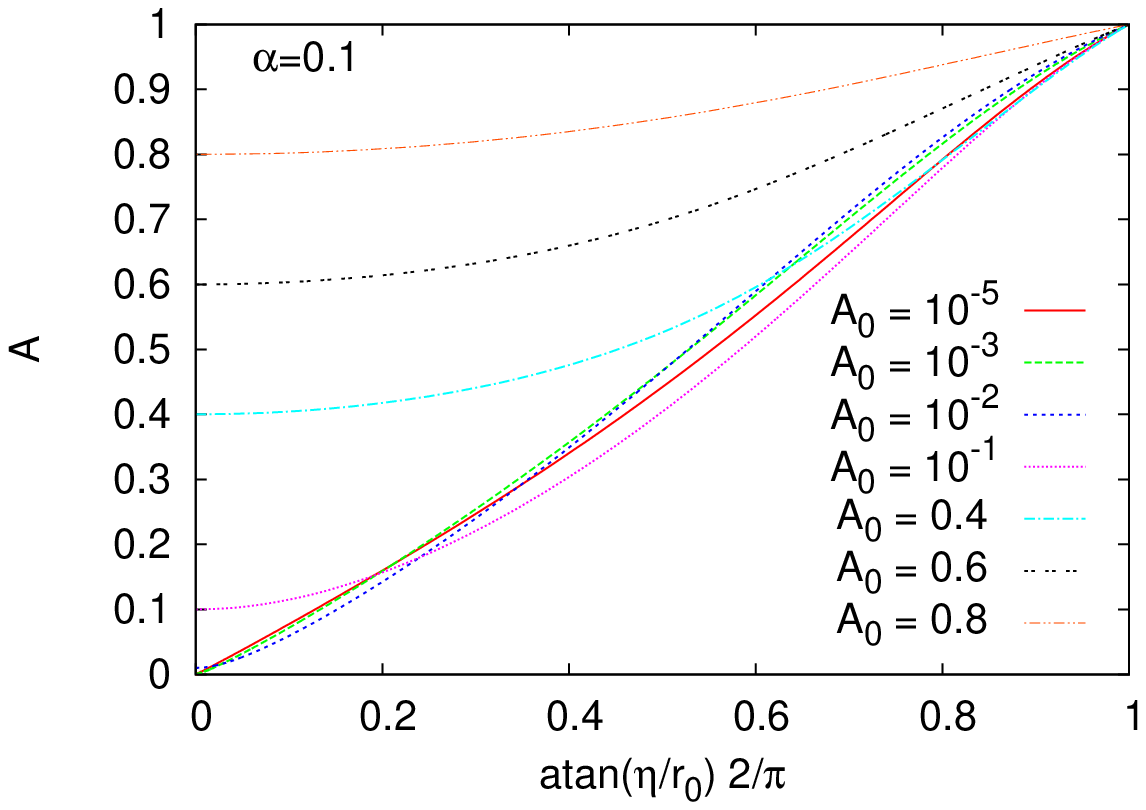}
\label{fig1a}
}
\subfigure[][]{\hspace{-0.5cm}
\includegraphics[height=.25\textheight, angle =0]{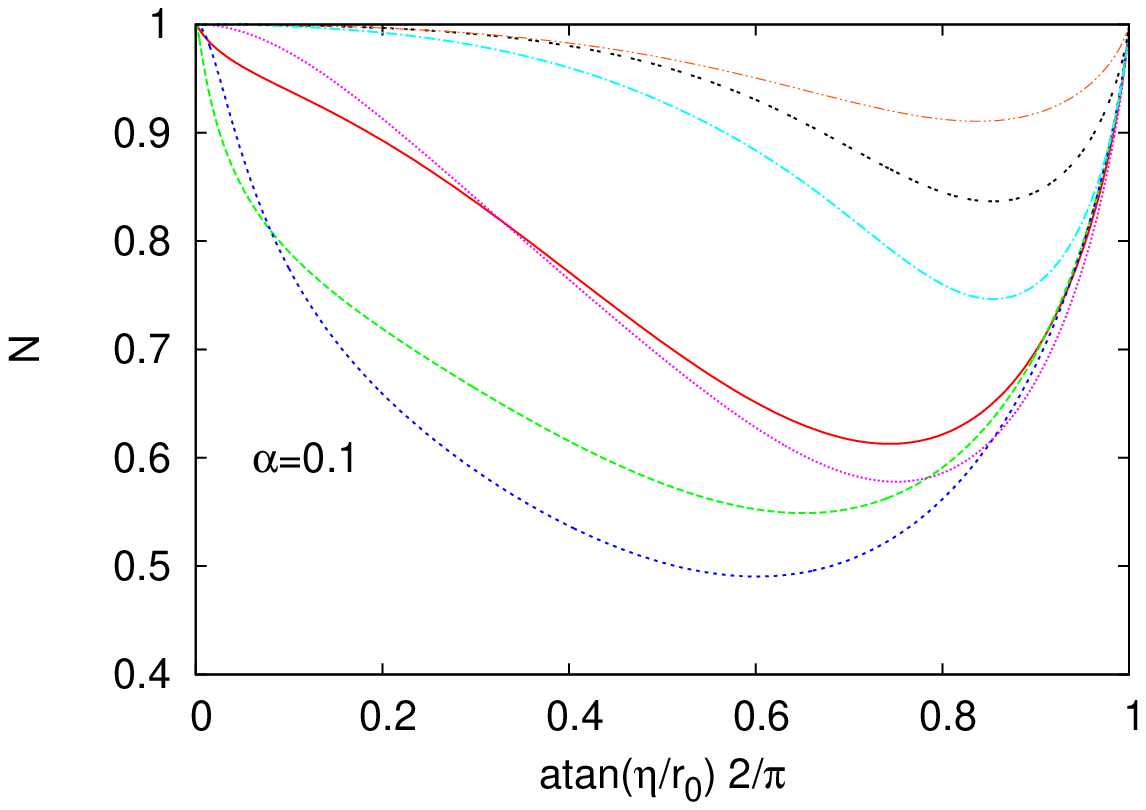}
\label{fig1b}
}
}
\mbox{\hspace{-0.5cm}
\subfigure[][]{\hspace{-1.0cm}
\includegraphics[height=.25\textheight, angle =0]{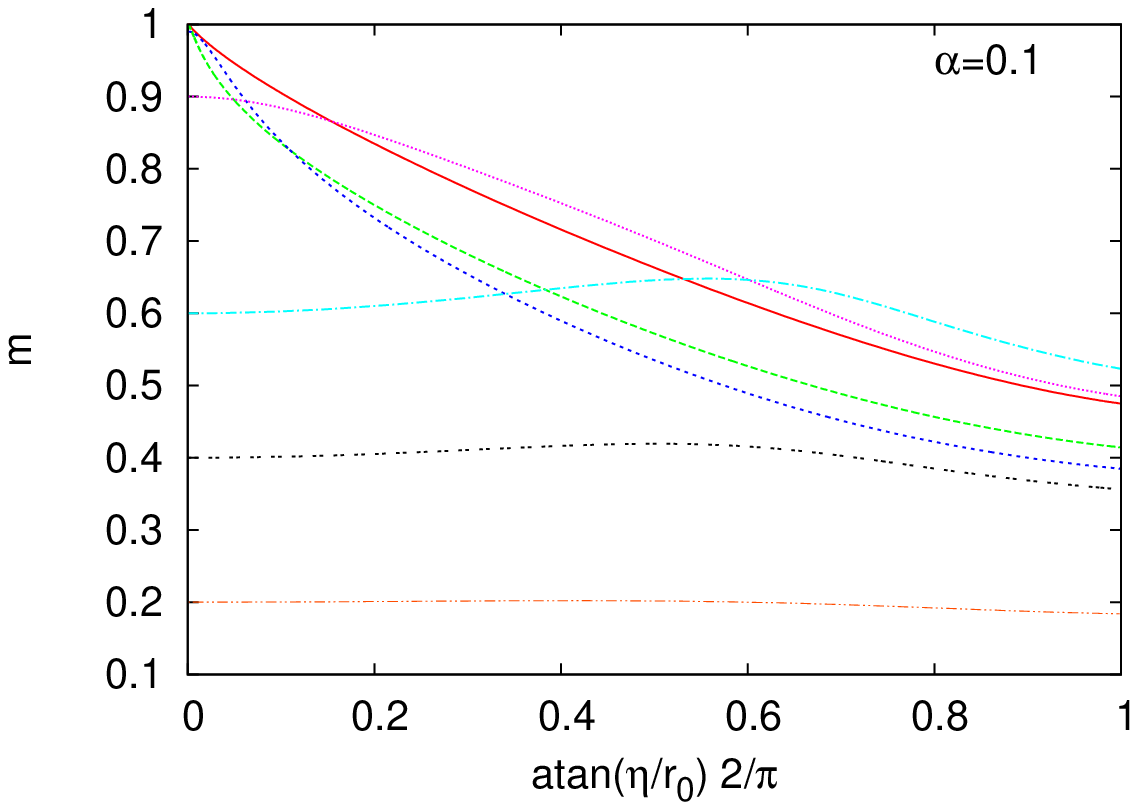}
\label{fig1c}
}
\subfigure[][]{\hspace{-0.5cm}
\includegraphics[height=.25\textheight, angle =0]{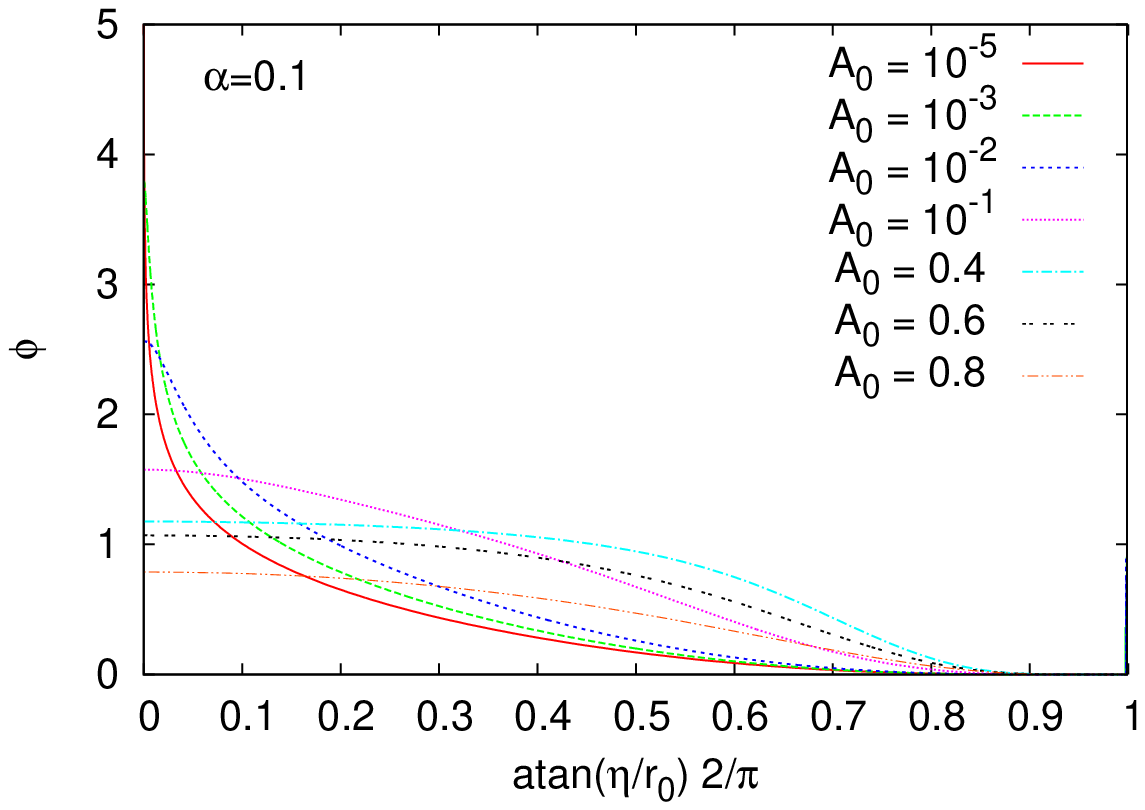}
\label{fig1d}
}
}
\end{center}
\vspace{-0.7cm}
\caption{
The metric functions $A$ (a) and $N$ (b),
the mass function $m$ (c) and the boson field function $\phi$ (d)
versus the compactified coordinate ${\rm atan}(\eta/r_0)$
for $\alpha=0.1$.
\label{fig1}
}
\end{figure}

We now consider the backreaction of the boson field on the metric
and thus the full set of coupled nonlinear field equations.
In the numerical calculations we replace the metric function
$R(\eta)$ in terms of the new metric function $N(\eta)$,
\begin{equation}
R^2(\eta) = N(\eta) \, (\eta^2 +r_0^2)  \ , 
\end{equation}
since in contrast to $R$ the new function $N$ is bounded
in the full interval $-\infty < \eta < \infty$.
The boundary conditions
$R(0)=r_0$ and $R'(0)=0$ then translate into the new conditions
$N(0)=1$ and  $N'(0)=0$, respectively.
We solve the ODEs numerically for the given set of
boundary conditions and a sequence of values 
for the parameters $\alpha$ and $r_0$.

Fixing the values of $\alpha$ and $r_0$, we then
obtain a family of boson star solutions with nontrivial topology.
We illustrate the solutions of such a family 
for $\alpha=0.1$ and $r_0=1$ in Fig.~\ref{fig1},
where the value of the metric function $A$ at the throat (or at the
equator)
$A_0=A(0)$, is varied.
At the same time the frequency $\omega$ of the solutions
changes, as discussed below.

We recall, that for the Ellis wormhole $A = 1$, and thus $A_0=1$. 
Then by decreasing $A_0$ from its limiting value
$A_0=1$ the family of boson star solutions with nontrivial topology
evolves from the limiting Ellis wormhole,
as seen in Fig.~\ref{fig1}.
In Fig.~\ref{fig1a}
the metric function $A(\eta)$ is exhibited as a function of the
compactified coordinate ${\rm atan}(\eta)$ for a sequence of values of $A_0$.
$A(\eta)$ is a monotonic function of $\eta$.
When $A_0$ tends to zero,
the function $A$ approaches a limiting function,
to be discussed later. 
$A_0$ cannot be decreased any further.

The metric function $N(\eta)$ of the same set of solutions is exhibited
in Fig.~\ref{fig1b}. Again, the starting solution is the Ellis solution, where
$N = 1$. As $A_0$ is decreased, the function $N$ develops
a minimum, which first deepens but then rises again.
%
We now introduce the mass function $m(\eta)$ via
\begin{equation}
A(\eta) = 1 - \frac{m(\eta)}{R(\eta)} \ .
\label{Amu}
\end{equation}
The mass $M$ of the solutions is obtained from its asymptotic value,
$m(\infty)=M$.
The mass function $m(\eta)$ is 
illustrated in Fig.~\ref{fig1c} for this set of solutions.
Identically zero for the limiting Ellis wormhole, 
the function $m(\eta)$ increases with decreasing $A_0$. 
However, this increase is monotonic only in the inner region
close to the throat. Here the mass function approaches
its maximal value $m(0) \to r_0$ as $A_0 \to 0$.
The asymptotic value $m(\infty)$ and thus the mass, on the other hand, 
is not changing monotonically for this family of solutions.

Finally, in Fig.~\ref{fig1d} we exhibit the boson field function
$\phi(x)$ for these solutions.
As expected the boson field function increases with decreasing $A_0$.
For the larger values of $A_0$ this behaviour is reminiscent
of the probe limit. In particular, also an extended inner plateau
appears. However,
as $A_0$ decreases further the plateau disappears again
and the function $\phi$ changes its character
and rises steeply towards the throat.
Note, that the value $\phi_0=\phi(0)$ can also be used to 
label the solutions of this family. 
In the limit $A_0 \to 0$, the boson field function
converges to a limiting function,
as discussed later.

\boldmath
\subsection{$\alpha$-Dependence}
\unboldmath

\begin{figure}[t!]
\begin{center}
\vspace{0.5cm}
\mbox{\hspace{-0.5cm}
\subfigure[][]{\hspace{-1.0cm}
\includegraphics[height=.25\textheight, angle =0]{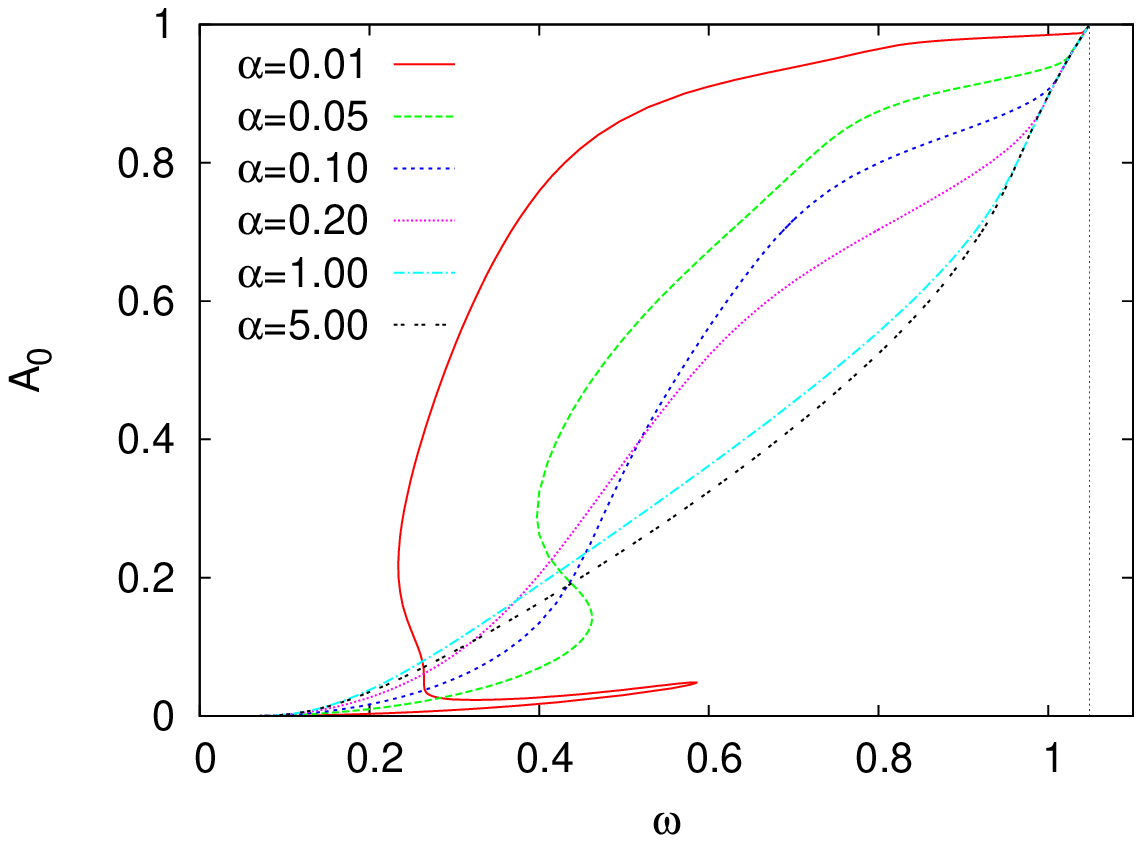}
\label{fig2a}
}
\subfigure[][]{\hspace{-0.5cm}
\includegraphics[height=.25\textheight, angle =0]{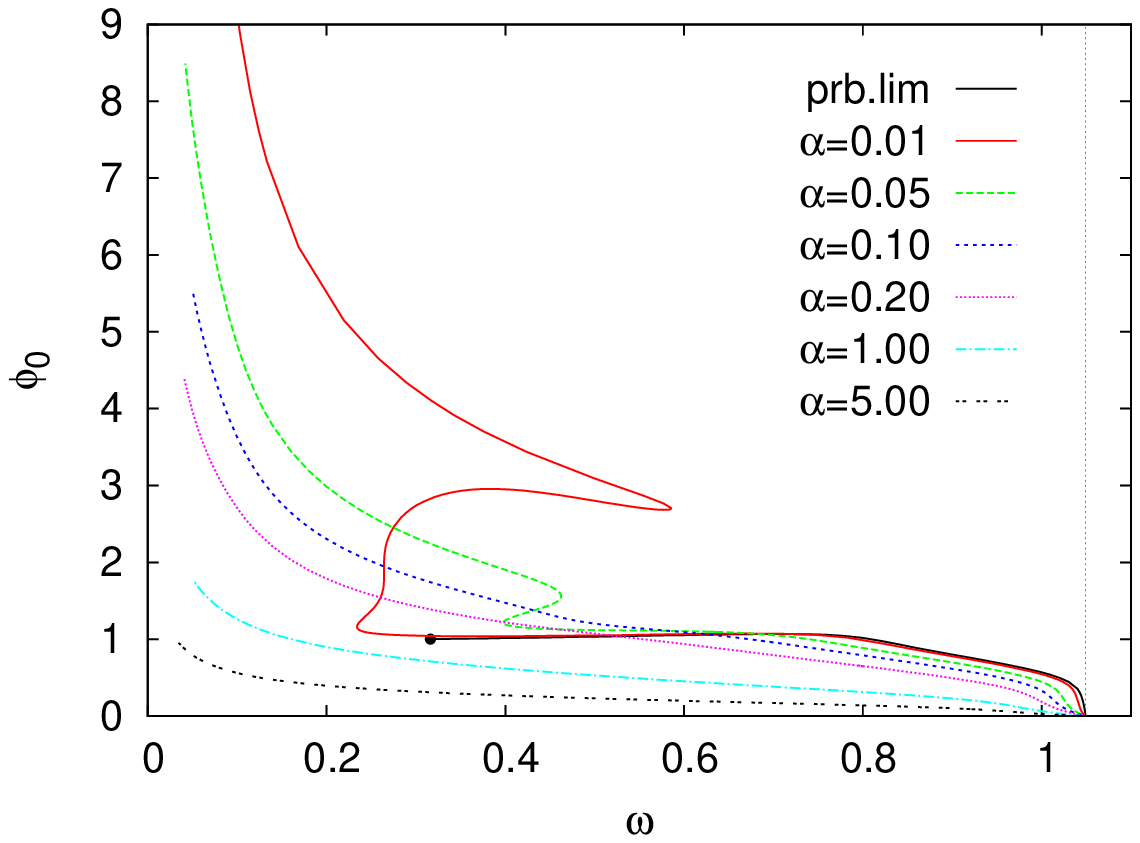}
\label{fig2b}
}
}
\mbox{\hspace{-0.5cm}
\subfigure[][]{\hspace{-1.0cm}
\includegraphics[height=.25\textheight, angle =0]{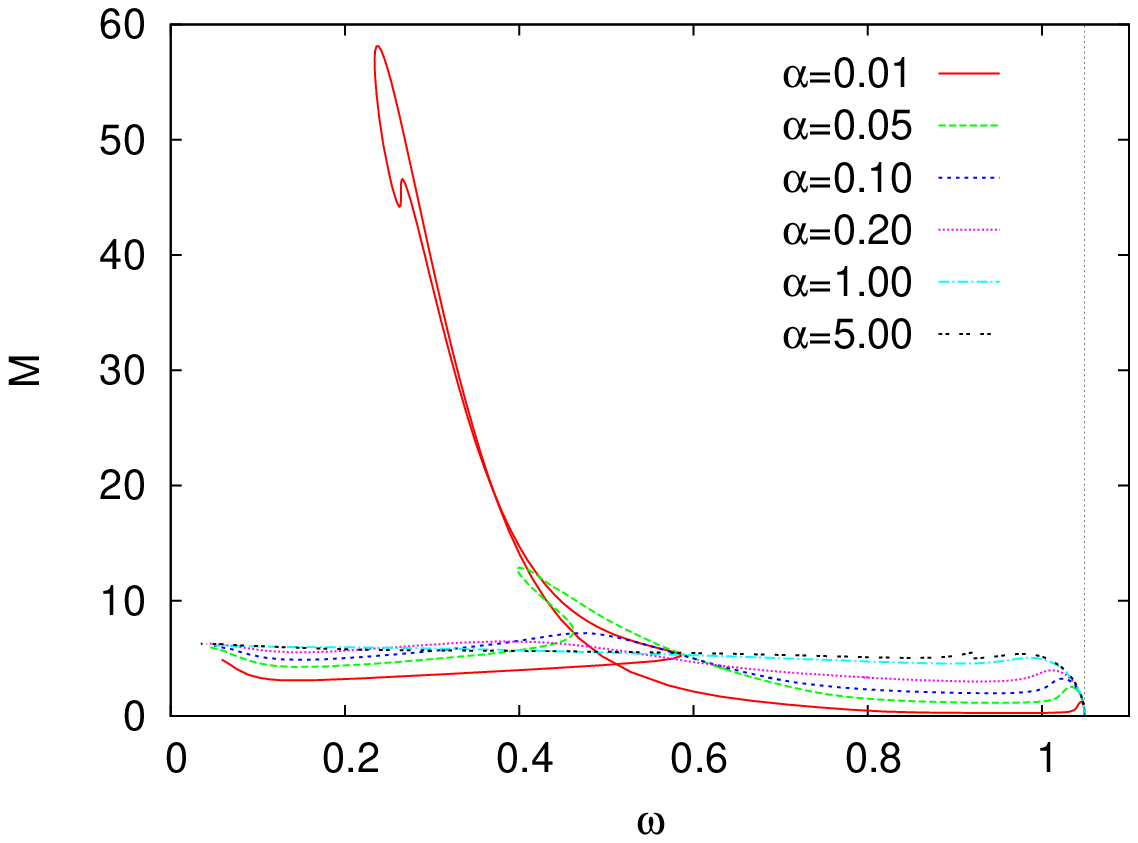}
\label{fig2c}
}
\subfigure[][]{\hspace{-0.5cm}
\includegraphics[height=.25\textheight, angle =0]{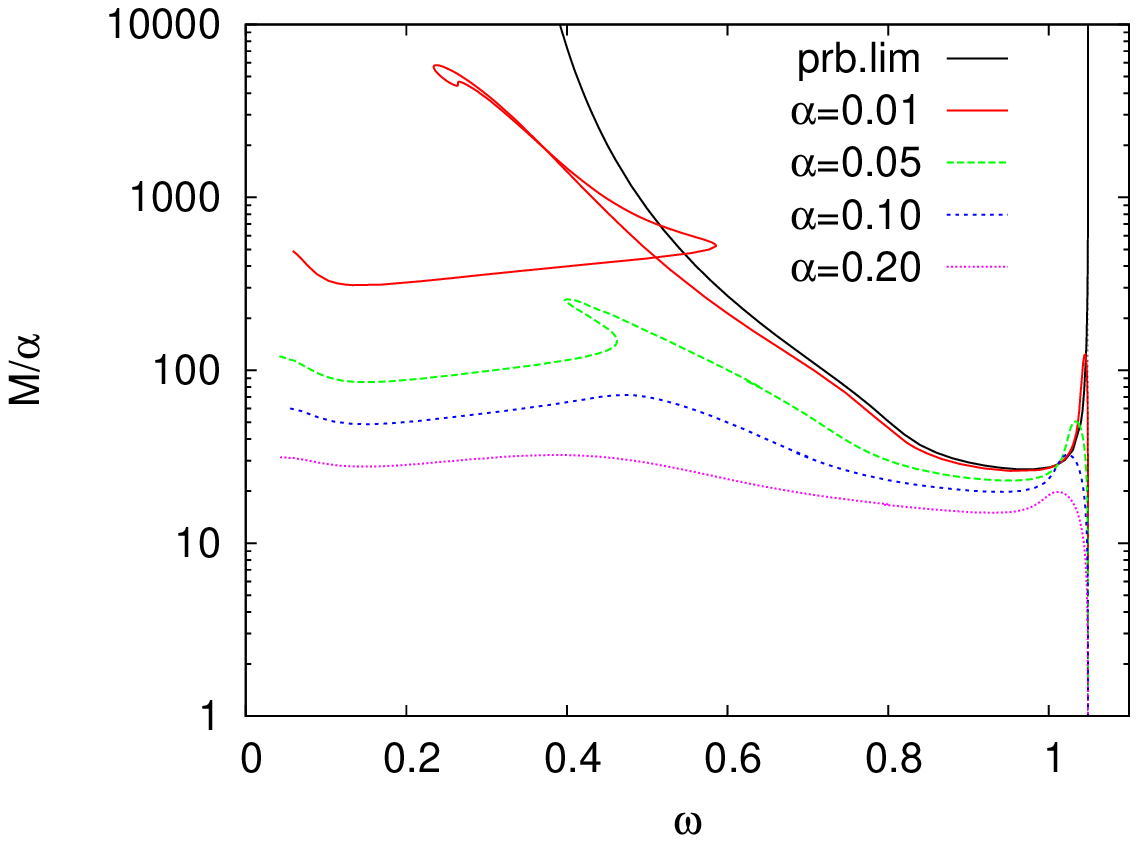}
\label{fig2d}
}
}
\mbox{\hspace{-0.5cm}
\subfigure[][]{\hspace{-1.0cm}
\includegraphics[height=.25\textheight, angle =0]{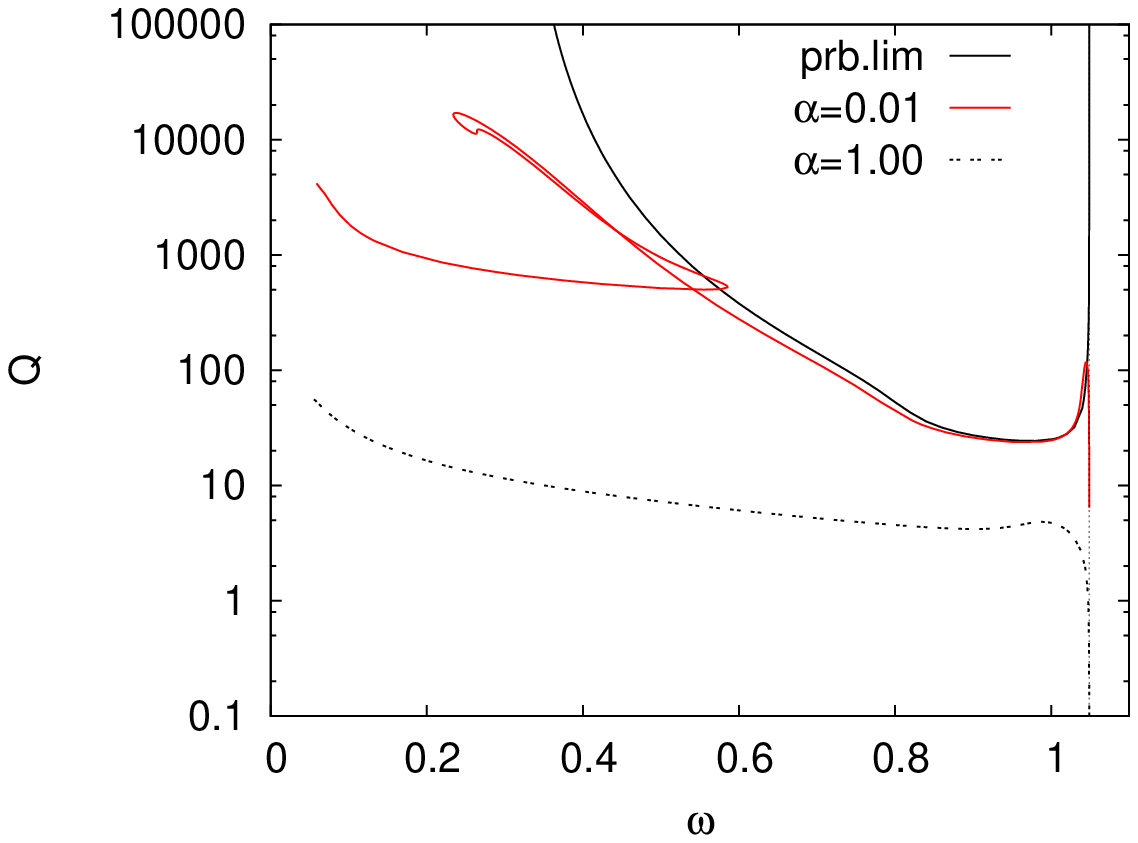}
\label{fig2e}
}
\subfigure[][]{\hspace{-0.5cm}
\includegraphics[height=.25\textheight, angle =0]{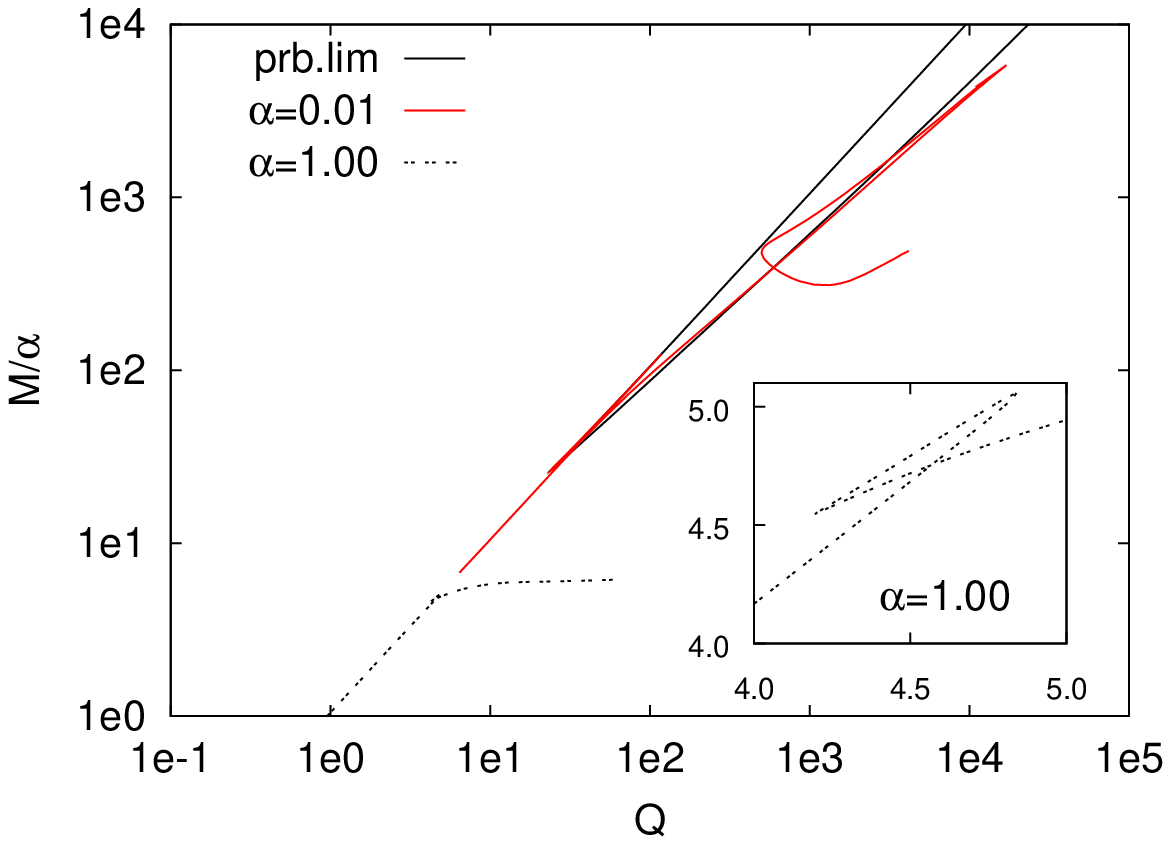}
\label{fig2f}
}
}
\end{center}
\vspace{-0.7cm}
\caption{
$\alpha$-dependence:
The values of the metric function 
$A_0$ (a) and the boson function $\phi_0$ (b)
at the throat (or equator),
the mass $M$ (c) 
the scaled mass $M/\alpha$ (d)
and the particle number $Q$ (e)
versus the boson frequency $\omega$.
The scaled mass $M/\alpha$ versus the particle number $Q$,
where the inset magnifies the cusp structure (f).
In (b), (d), (e) and (f) the solid black curves represent the probe limit.
The probe limit of (a) is trivial, $A_0=1$.
The thin vertical line indicates $\omega_{\rm max}$.
\label{fig2}
}
\end{figure}

The family of $\alpha=0.1$
solutions can also be considered as labelled by the frequency
$\omega$ of the boson field, as done above in the probe limit. 
The frequency dependence
has been studied widely for boson stars
\cite{Friedberg:1986tq}, where the coupling to gravity
leads to a very different behaviour for large and for
small frequencies as compared to $Q$-balls.
The maximal value $\omega_{\rm max}$ is retained,
when gravity is coupled. However, the value $\omega_{\rm min}$
no longer represents the minimal frequency for boson stars
for generic values of $\alpha$.
In particular, for small values of the frequency
the families of boson stars feature spirals,
when the mass or the particle number are considered
versus the frequency.
But with decreasing $\alpha$, the $Q$-ball limit
is approached in an increasing interval of $\omega$.

We illustrate the dependence of the
families of solutions on the coupling constant $\alpha$ in Fig.~\ref{fig2}.
Besides the solutions with $\alpha=0.1$, considered above, here 
also solutions for larger values and smaller values of $\alpha$ are shown.
For comparison also the probe limit is included in the figure.

We exhibit in Fig.~\ref{fig2a} the value of $A_0$ for these
families of solutions versus the frequency $\omega$.
In the probe limit $A_0=1$. 
As $\alpha$ is increased from zero,
and the backreaction of the boson field on the metric is taken into account,
$A_0$ starts to deviate from the probe limit. 
For small $\alpha$ this deviation is small in a large region of $\omega$
in its upper allowed range.
For smaller values of $\omega$, however, $A_0$ decreases dramatically,
and then tends to zero.
For larger values of the coupling constant
$\alpha$ the decrease of $A_0$ becomes more uniform.

The value $\phi_0$ is exhibited in Fig.~\ref{fig2b} for the same
families of solutions.
Again, we observe, that for $\alpha \to 0$ the respective curve
of the probe limit is approached in most of the interval
allowed in the probe limit, $\omega_{\rm min} < \omega < \omega_{\rm max}$.
However, close to $\omega_{\rm min}$ we observe
significant deviations, since the solutions continue 
to values of $\omega$ 
beyond $\omega_{\rm min}$.
In particular, $\phi_0$ starts to increase strongly
for small values of $\alpha$ and $\omega$.

The mass $M$ and the scaled mass $M/\alpha$
are shown versus the frequency $\omega$ 
in Fig.~\ref{fig2c} and \ref{fig2d},
respectively.
We need to scale the mass in order to compare with the probe limit.
We observe, that the boson star solutions with nontrivial
topology tend towards their probe limit as $\alpha \to 0$,
in the interval $\omega_{\rm min} < \omega < \omega_{\rm max}$.
As for ordinary boson stars the mass 
rises from zero at $\omega_{\rm max}$ and approaches a local maximum.
Subsequently, it reaches a local minimum from where it rises again.
For small values of $\omega$ the mass converges
to a finite common value, 
independent of $\alpha$.

As seen in Fig.~\ref{fig2e},
the particle number $Q$ follows the mass closely
in most of the $\omega$-interval.
Thus for $\alpha \to 0$,
$Q$ approaches the probe limit analogously.
However, its $\omega$-dependence differs for small values of $\omega$.
In particular, the particle number keeps increasing
with decreasing $\omega$, as indicated in Fig.~\ref{fig2f}.


\subsection{Wormhole Geometries}

\begin{figure}[t!]
\begin{center}
\vspace{0.5cm}
\mbox{\hspace{-0.5cm}
\subfigure[][]{\hspace{-1.0cm}
\includegraphics[height=.25\textheight, angle =0]{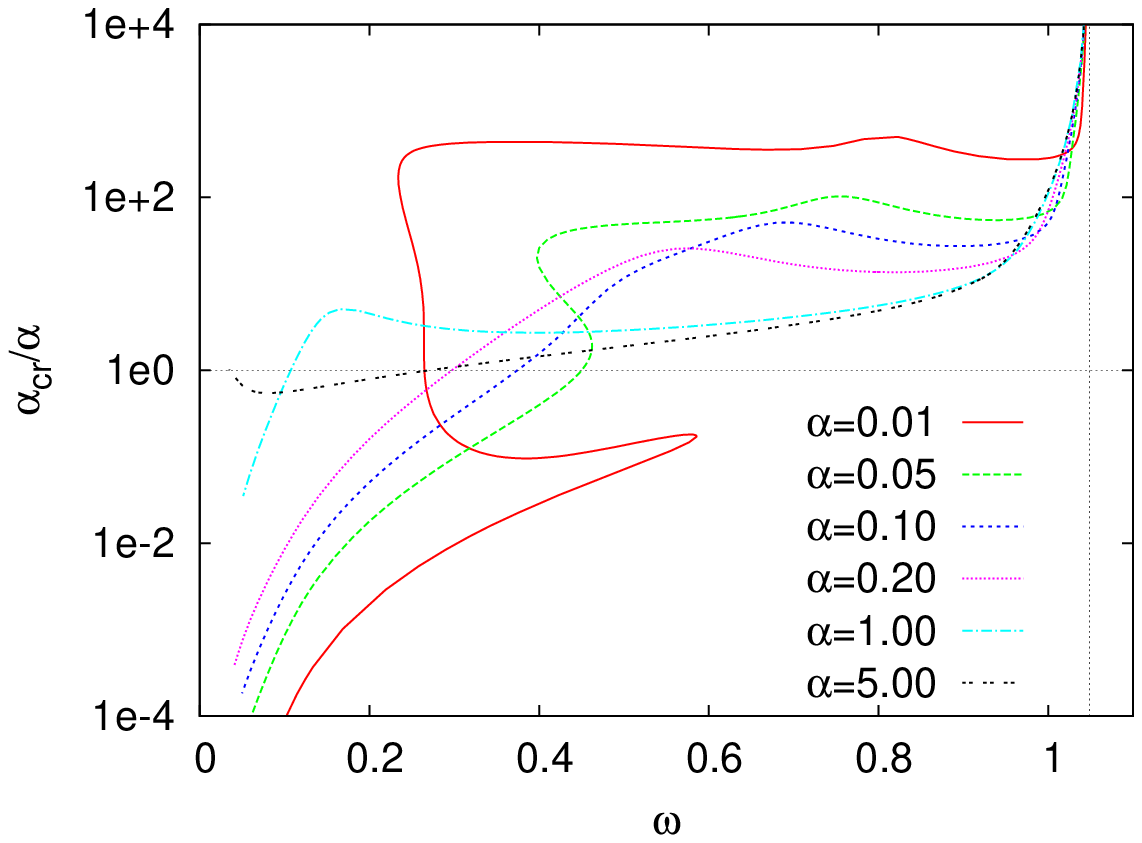}
\label{fig3a}
}
\subfigure[][]{\hspace{-0.5cm}
\includegraphics[height=.25\textheight, angle =0]{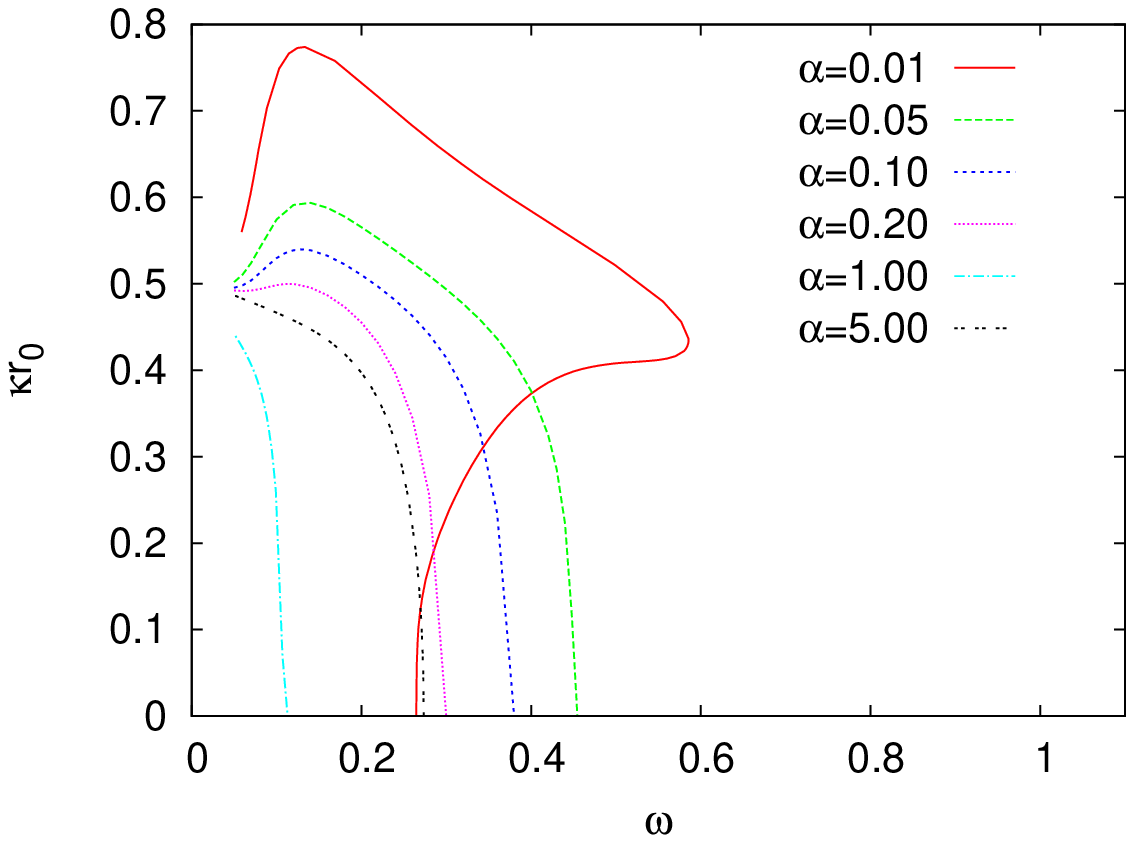}
\label{fig3b}
}
}
\end{center}
\vspace{-0.7cm}
\caption{
Single or double throat: The ratio of the critical value $\alpha_{\rm cr}$
to $\alpha$ (a)
and the surface gravity $\kappa$ (b)
versus the boson frequency $\omega$
for several values of $\alpha$.
\label{fig3}
}
\end{figure}

We now turn to the geometry of the wormholes within the boson stars.
Here our first quest is to find out whether the wormholes possess a single
throat or a double throat,
where the latter will arise because of the backreaction of the boson
field on the metric.
For this purpose we evaluate the critical value $\alpha_{\rm cr}$,
defined in Eq.~\ref{acrit}.
We exhibit the ratio $\alpha_{\rm cr}/\alpha $ versus the frequency $\omega$
for several families of solutions in Fig.~\ref{fig3a}.
As long as $\alpha_{\rm cr}/\alpha > 1$ the solutions possess only
a single throat.

We show the surface gravity $\kappa$ for
the same sets of solutions in
Fig.~\ref{fig3b}. Here $\kappa=0$ signals that the solutions
possess only a single throat, whereas finite values of $\kappa$
imply the presence of an equator and a double throat.

\begin{figure}[t!]
\begin{center}
\vspace{0.5cm}
\mbox{\hspace{-0.5cm}
\subfigure[][]{\hspace{-1.0cm}
\includegraphics[height=.25\textheight, angle =0]{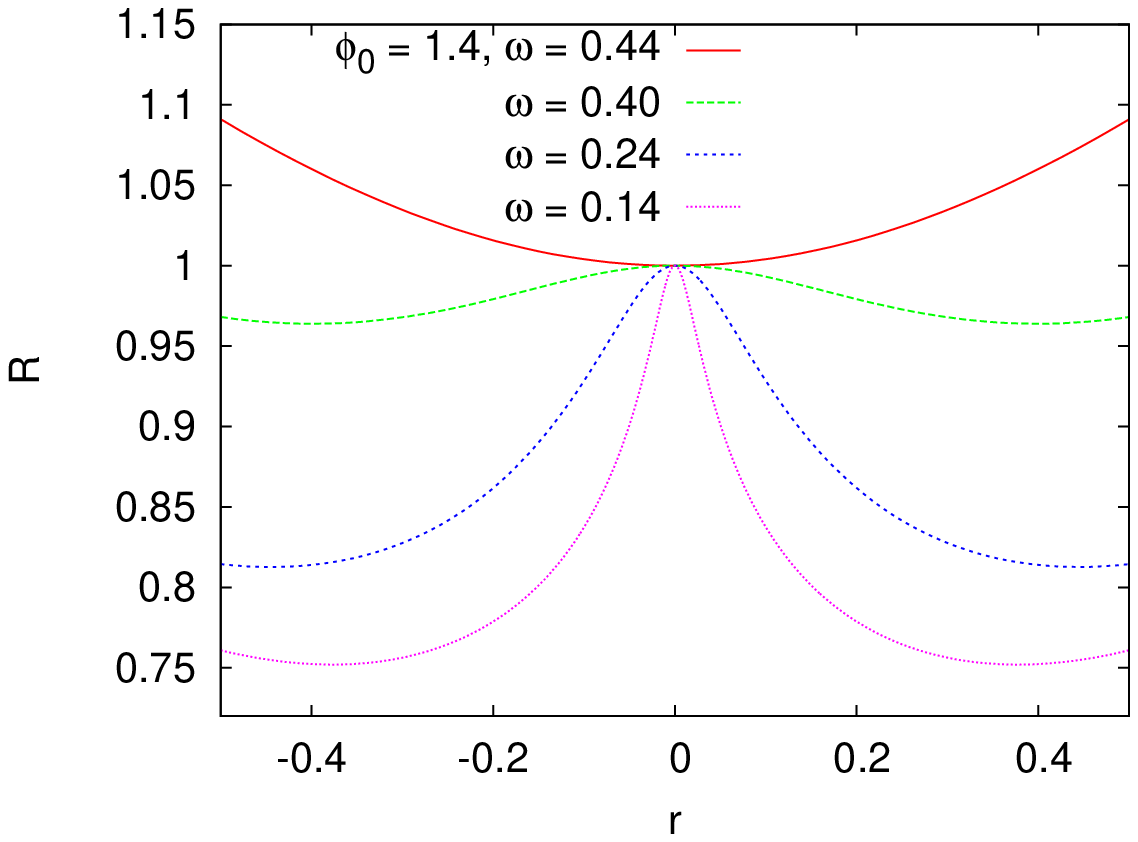}
\label{fig4a}
}
\subfigure[][]{\hspace{-0.5cm}
\includegraphics[height=.25\textheight, angle =0]{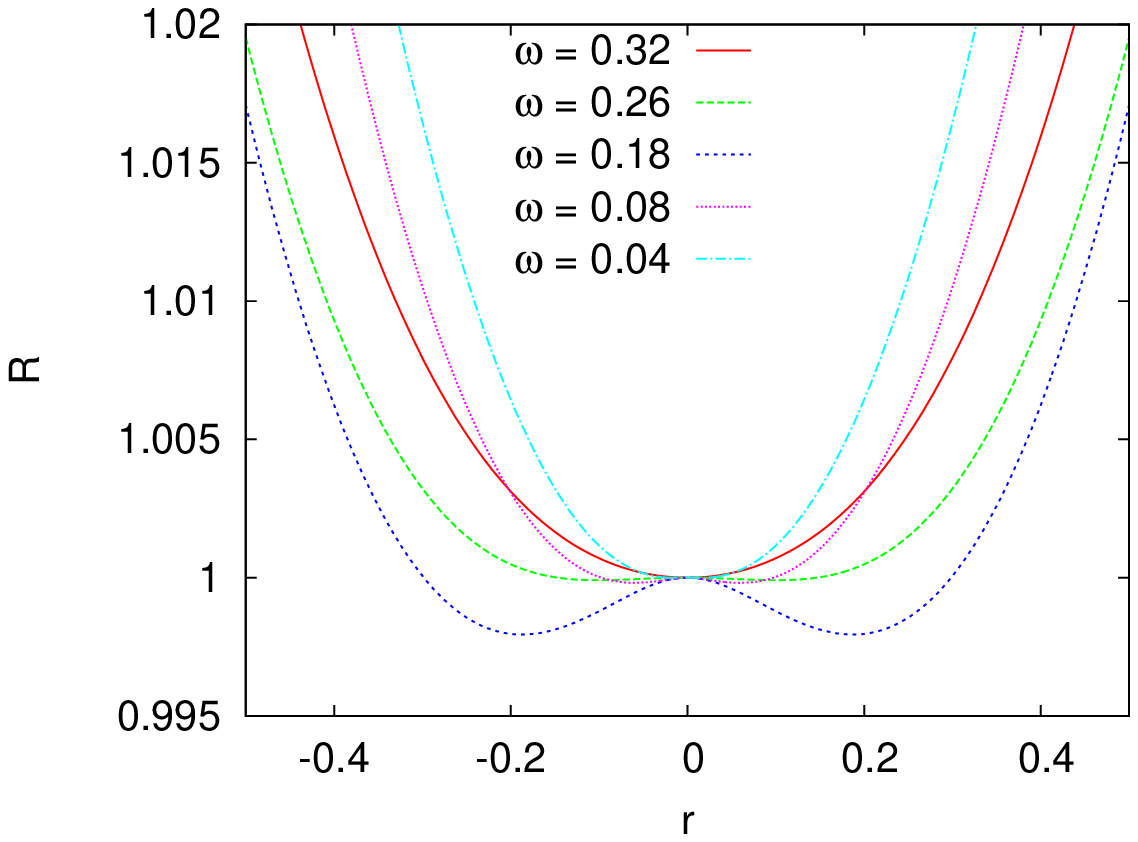}
\label{fig4b}
}
}
\end{center}
\vspace{-0.7cm}
\caption{
Throat geometry:
The metric function $R$ versus the radial coordinate $\eta$
for a sequence of solutions with various values
of the frequency $\omega$ and coupling constant
$\alpha = 0.05$ (a) and $\alpha = 5$ (b).
\label{fig4}
}
\end{figure}

In Fig.~\ref{fig4} we exhibit the metric function $R$
for a number of solutions
for two values of the coupling constant, $\alpha=0.05$
and $\alpha=5$.
For $\omega$ close to $\omega_{\rm max}$
the solutions always possess only a single throat,
since here $\alpha < \alpha_{\rm cr}$.
For families with large $\alpha$, this condition is
soon violated, when $\omega$ decreases,
and the solutions exhibit a double throat.
Then $R$ possesses a local maximum at $\eta=0$ 
and two minima located symmetrically on each side of the local maximum. 
For families with small $\alpha$, on the other hand,
solutions with a double throat appear only for
much smaller values of the frequency.
However, close to the limiting solutions discussed below,
the solutions always exhibit a double throat.

\begin{figure}[t!]
\begin{center}
\vspace{0.5cm}
\mbox{\hspace{-0.5cm}
\subfigure[][]{\hspace{-1.0cm}
\includegraphics[height=.25\textheight, angle =0]{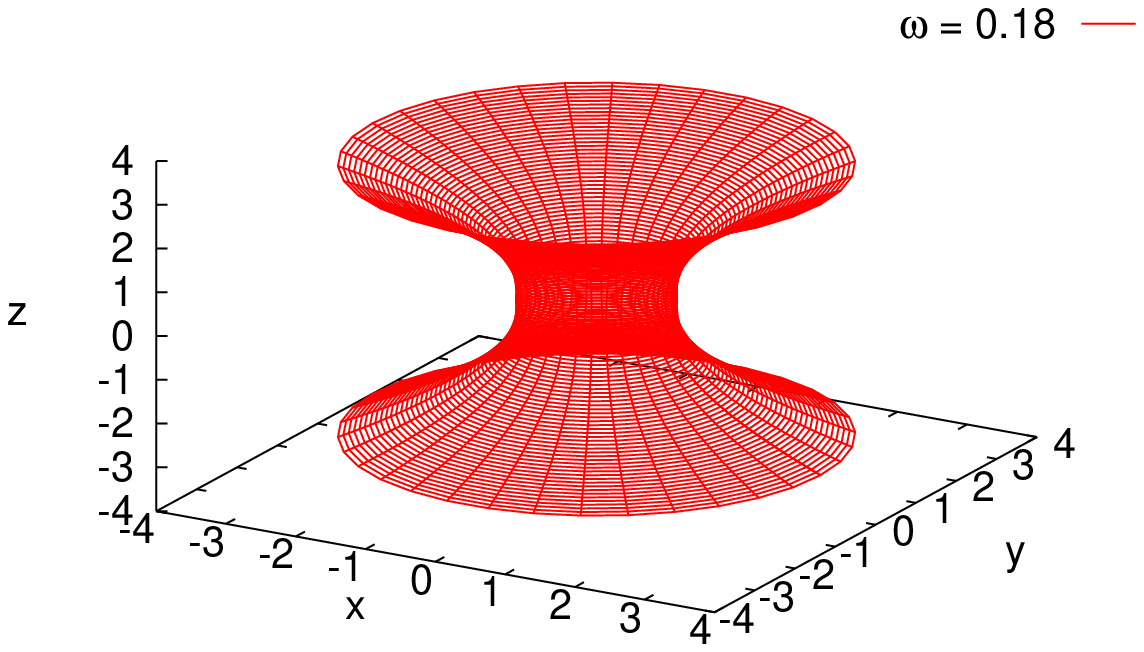}
\label{fig5a}
}
\subfigure[][]{\hspace{-0.5cm}
\includegraphics[height=.25\textheight, angle =0]{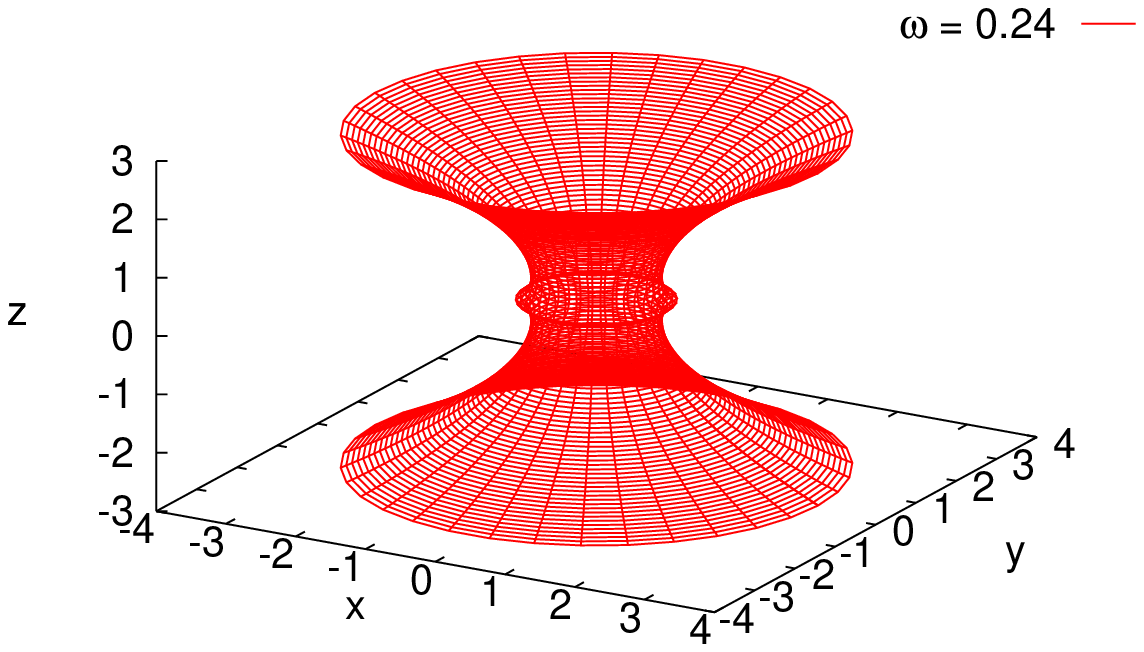}
\label{fig5b}
}
}
\end{center}
\vspace{-0.7cm}
\caption{
Throat geometry:
Three dimensional view of the isometric embedding for 
$\alpha=5$, $\omega=0.18$ (a) and
$\alpha=0.05$, $\omega=0.24$ (b).
\label{fig5}
}
\end{figure}

We visualize the geometry of a spatial hypersurface 
of the wormhole space-times in Fig.~\ref{fig5},
where we display an isometric embedding of the equatorial plane
$\theta= \pi/2$.
For the embedding the parametric representation
\begin{equation}
\rho(\eta) = R(\eta) \ , \ \ \ \ 
z(\eta)= \int_0^\eta\sqrt{1-R'^2} \ d\eta' \ ,
\label{embedd}
\end{equation}
is employed.
The radius $R$ has a single minimum at $z=0$, when
$\alpha< \alpha_{\rm cr}$. 
This minimum is seen as the waist in Fig.~\ref{fig5a}. 
At $\alpha=\alpha_{\rm cr}$ the minimum becomes degenerate.
For $\alpha> \alpha_{\rm cr}$, finally,
$z=0$ turns into a (local) maximum.
This corresponds to the belly in Fig.~\ref{fig5b}.
We do not find solutions with three or more throats here.

\boldmath
\subsection{Limit $A_0 \to 0$}
\unboldmath

\begin{figure}[t!]
\begin{center}
\vspace{0.5cm}
\mbox{\hspace{-0.5cm}
\subfigure[][]{\hspace{-1.0cm}
\includegraphics[height=.25\textheight, angle =0]{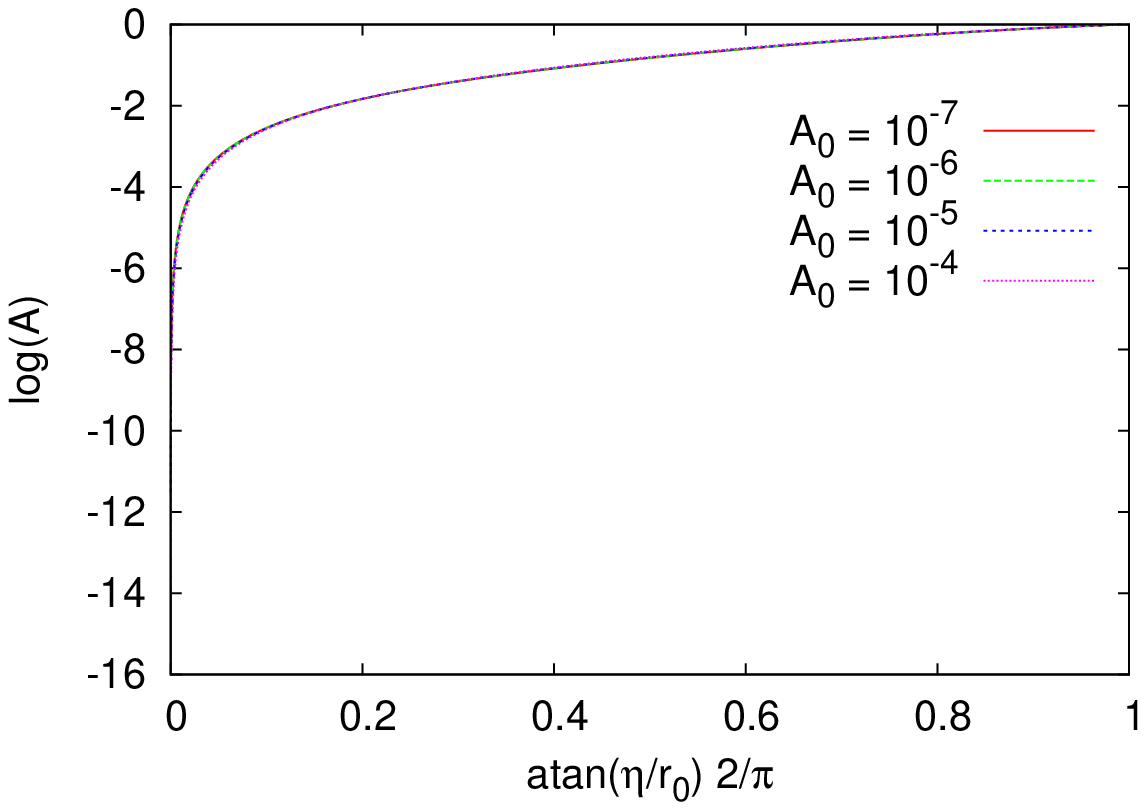}
\label{fig6a}
}
\subfigure[][]{\hspace{-0.5cm}
\includegraphics[height=.25\textheight, angle =0]{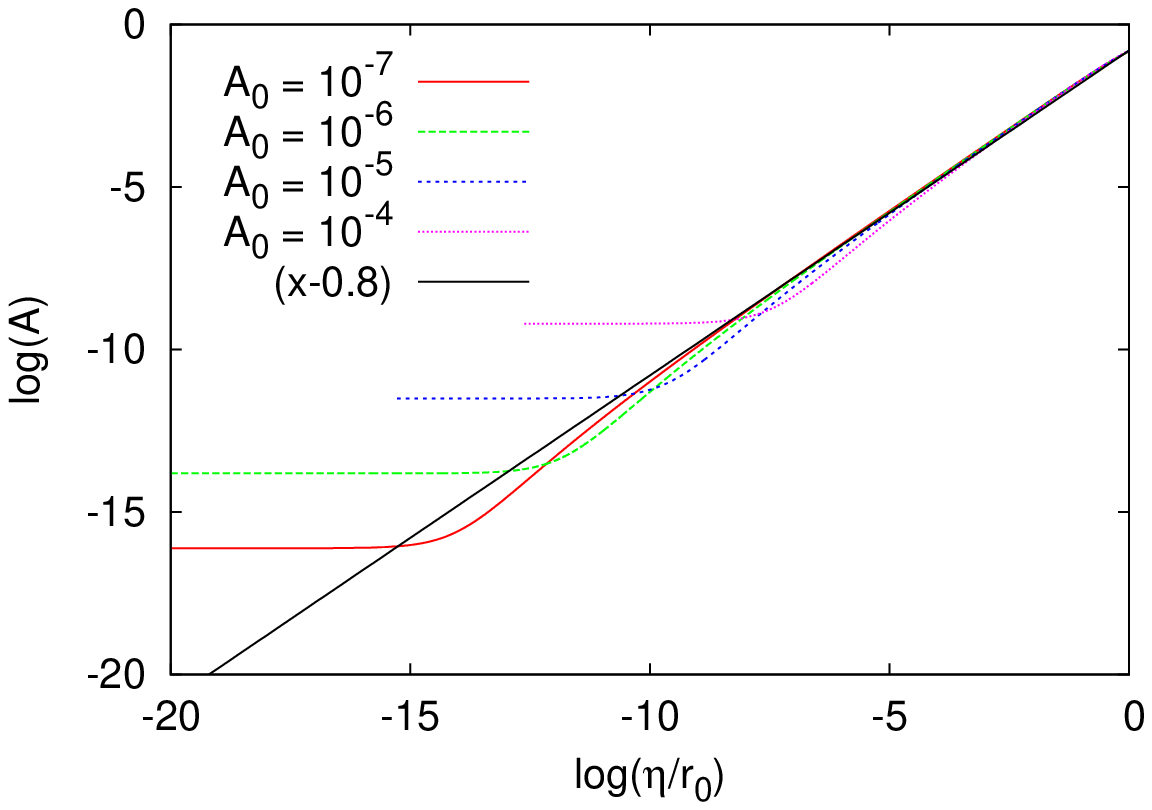}
\label{fig6b}
}
}
\mbox{\hspace{-0.5cm}
\subfigure[][]{\hspace{-1.0cm}
\includegraphics[height=.25\textheight, angle =0]{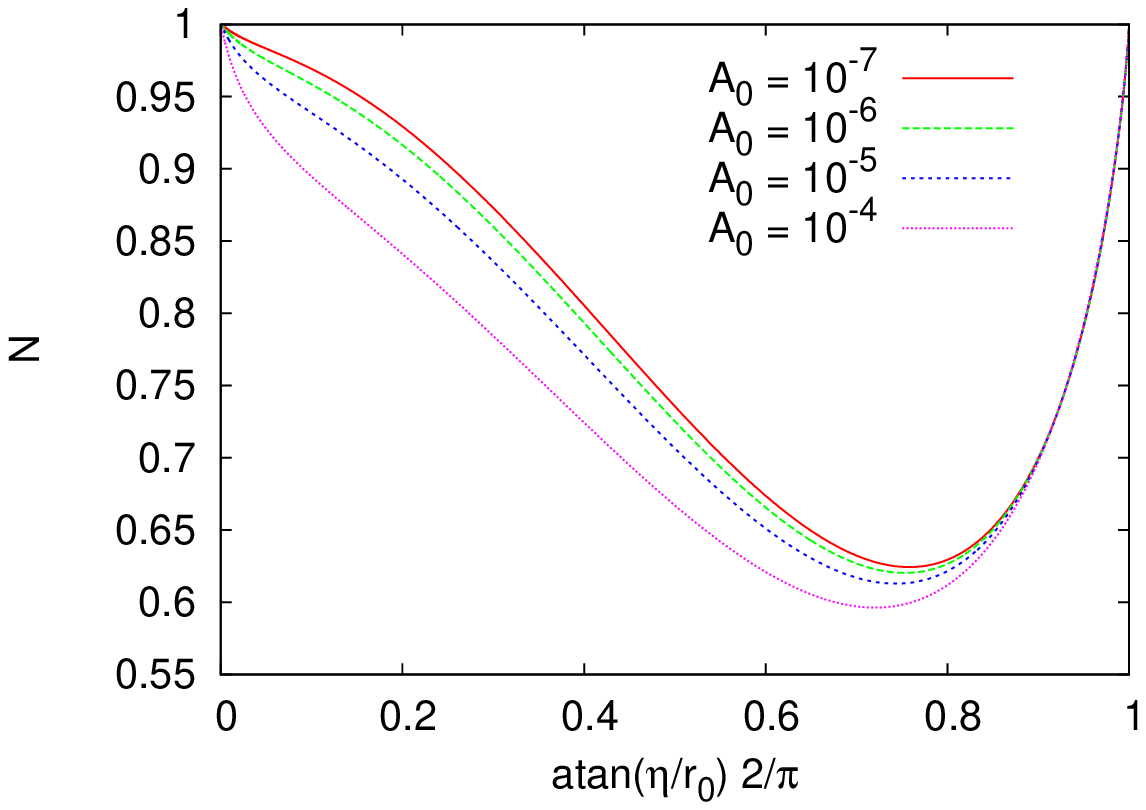}
\label{fig6c}
}
\subfigure[][]{\hspace{-0.5cm}
\includegraphics[height=.25\textheight, angle =0]{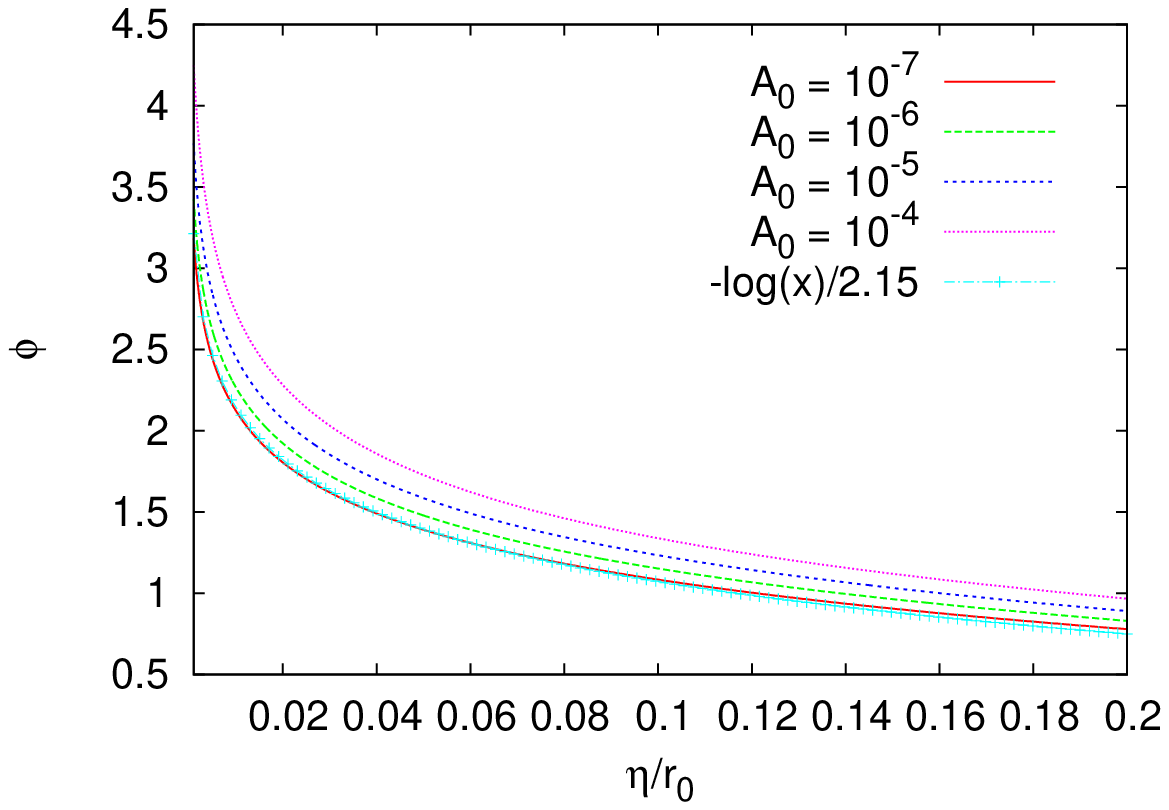}
\label{fig6d}
}
}
\mbox{\hspace{-0.5cm}
\subfigure[][]{\hspace{-1.0cm}
\includegraphics[height=.25\textheight, angle =0]{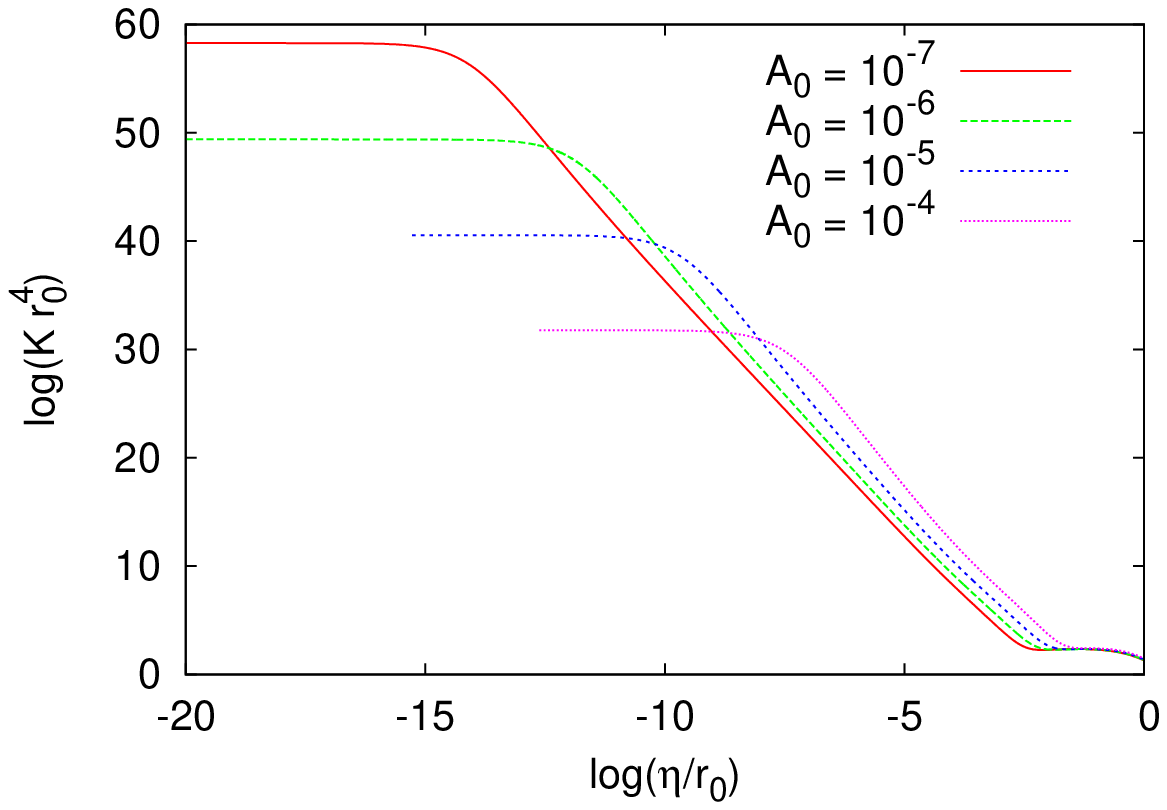}
\label{fig6e}
}
\subfigure[][]{\hspace{-0.5cm}
\includegraphics[height=.25\textheight, angle =0]{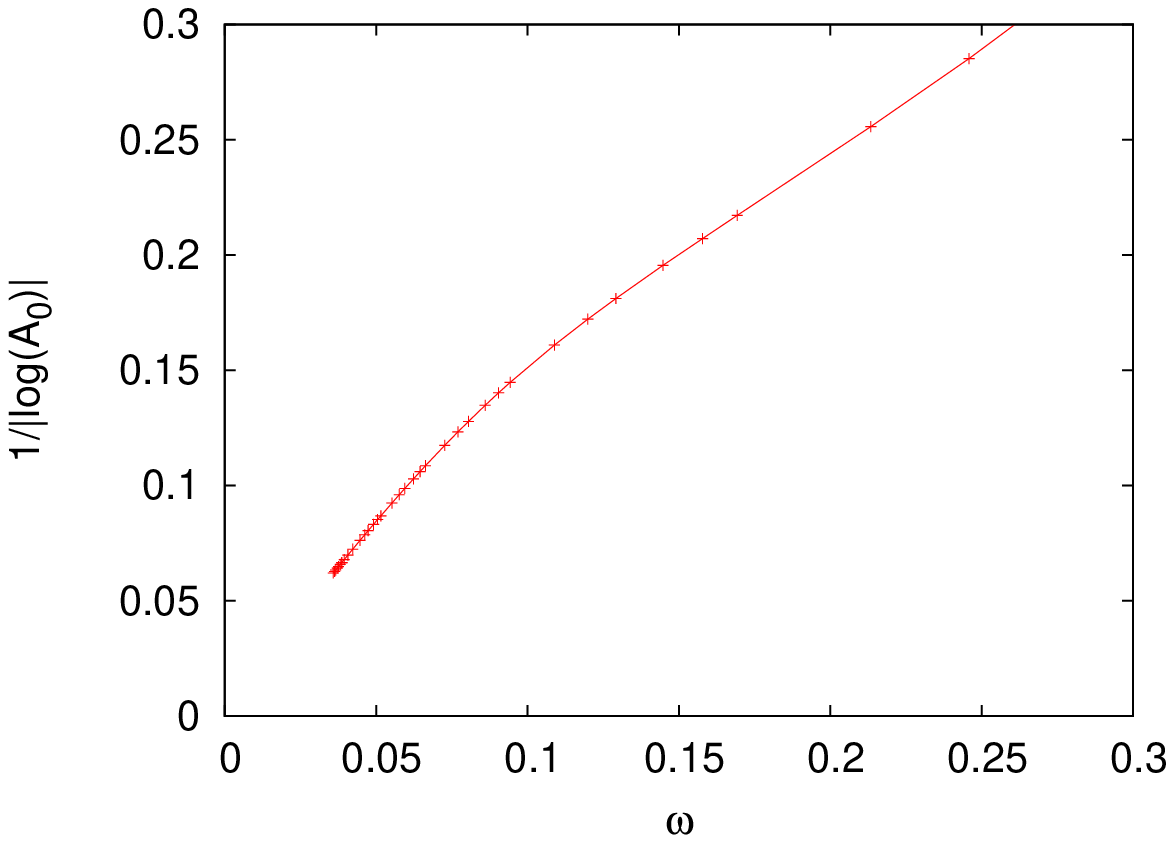}
\label{fig6f}
}
}
\end{center}
\vspace{-0.7cm}
\caption{
Limiting behaviour of the solutions for $\omega \to 0$:
The logarithm of the metric function $A$ versus the compactified coordinate 
${\rm atan}\, \eta$ (a),
a zoom of ${\rm ln}\, A$ versus ${\rm ln}\, \eta$ with the (approximate)
limiting behavior close to $\eta=0$ (b),
the metric function $N$ versus ${\rm atan}\, \eta$ (c),
a zoom of the boson field function $\phi$ versus $\eta$
with the (approximate) limiting behavior close to $\eta=0$ (d),
and the Kretschmann scalar $K$ versus ${\rm ln}\, \eta$ (e)
for several values of $A_0$ and $\alpha=0.1$.
The function $1/{\rm ln}\, A_0$, where $A_0=A(0)$, versus
the boson frequency $\omega$ for $\alpha=0.1$.
\label{fig6}
}
\end{figure}

Let us now address the limiting solution
reached, when $\omega \to 0$.
We demonstrate the limiting behaviour in Fig.~\ref{fig6}.
For this purpose we select a sequence of solutions
with decreasing values of $A_0=A(0)$ for $\alpha=0.1$.  
In Fig.~\ref{fig6a} we see, that the metric function $A$
is converging fast to its limiting function.
The small deviations are highlighted in Fig.~\ref{fig6b},
where we zoom into the region close to the equator, $\eta=0$.
The limiting function in this region is approximately described by
${\rm ln}\, A = {\rm ln}\, \eta -0.8$.
We note, that this limiting function is independent of $\alpha$.

In Fig.~\ref{fig6c} we exhibit the limiting behavior of the metric
function $N$. Here full convergence has only been achieved for large
$\eta$. 
When the function $R$ is considered instead of $N$, one notices,
that $R''(0)$ diverges in the limit.
The boson field function $\phi$ also converges to a limiting function.
This is seen in Fig.~\ref{fig6d}, where we zoom into the region close to the equator
again. 
In this region the limiting boson function is approximately described by
$\phi={\rm ln}\, \eta/2.15$.

Thus, as $\omega \to 0$, the solutions tend to a limiting
solution with a singular behavior
at its core.
By evaluating the Kretschmann scalar 
$K=R_{\mu\nu\alpha\beta} R^{\mu\nu\alpha\beta}$ of the solutions,
we see that
this singular behavior can be attributed to a curvature
singularity. 
The Kretschmann scalar of the same set of solutions
is shown in Fig.~\ref{fig6e}. 
Clearly, for $A_0 \to 0$ the Kretschmann scalar
will diverge at $\eta=0$,
confirming the presence of a curvature
singularity in the limiting solution. 

Finally, the dependence of $A_0$ on the boson frequency $\omega$
is demonstrated in Fig.~\ref{fig6f}. Here we display
the function $1/{\rm ln}\, A_0$, to resolve the limiting
behavior. 
The figure clearly indicates, that the singular limit should be reached
for $\omega \to 0$.
But the emerging singularity prevents
numerical calculations closer to the limit.

\subsection{Astrophysical Properties}

\begin{figure}[t!]
\begin{center}
\vspace{0.5cm}
\mbox{\hspace{-0.5cm}
\subfigure[][]{\hspace{-1.0cm}
\includegraphics[height=.25\textheight, angle =0]{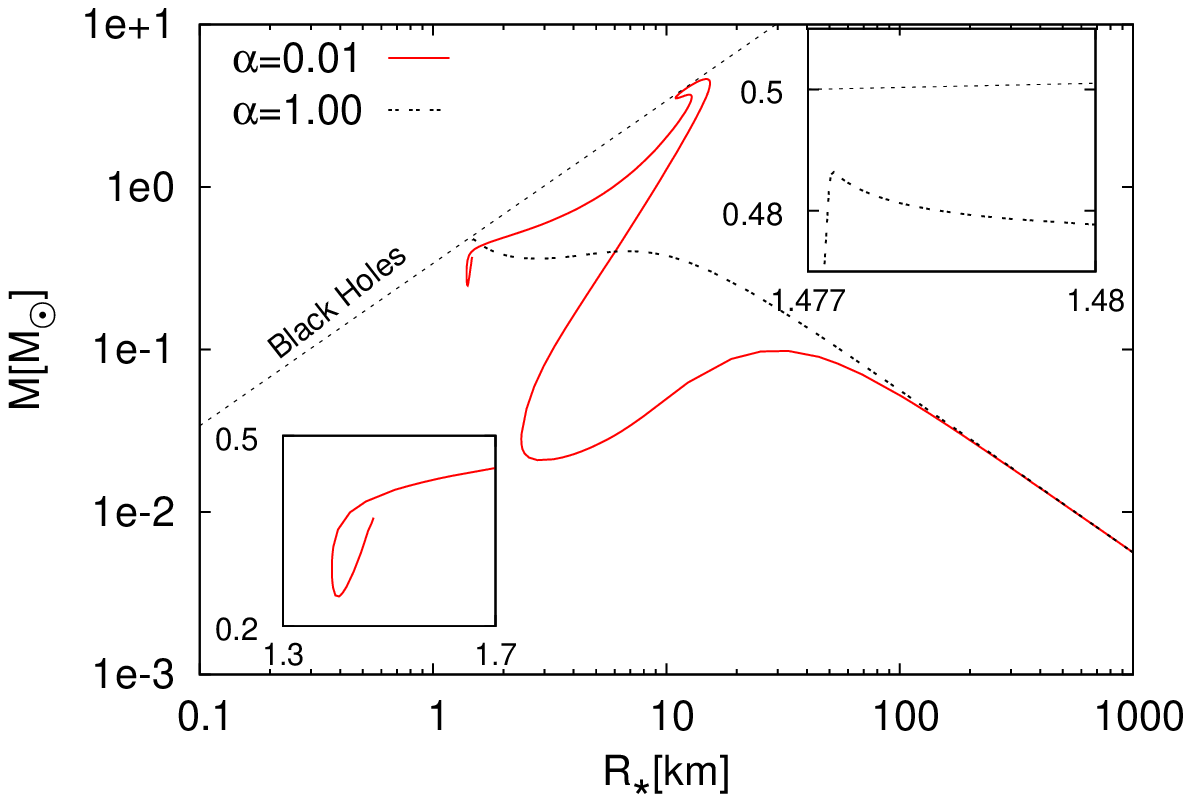}
\label{fig7a}
}
\subfigure[][]{\hspace{-0.5cm}
\includegraphics[height=.25\textheight, angle =0]{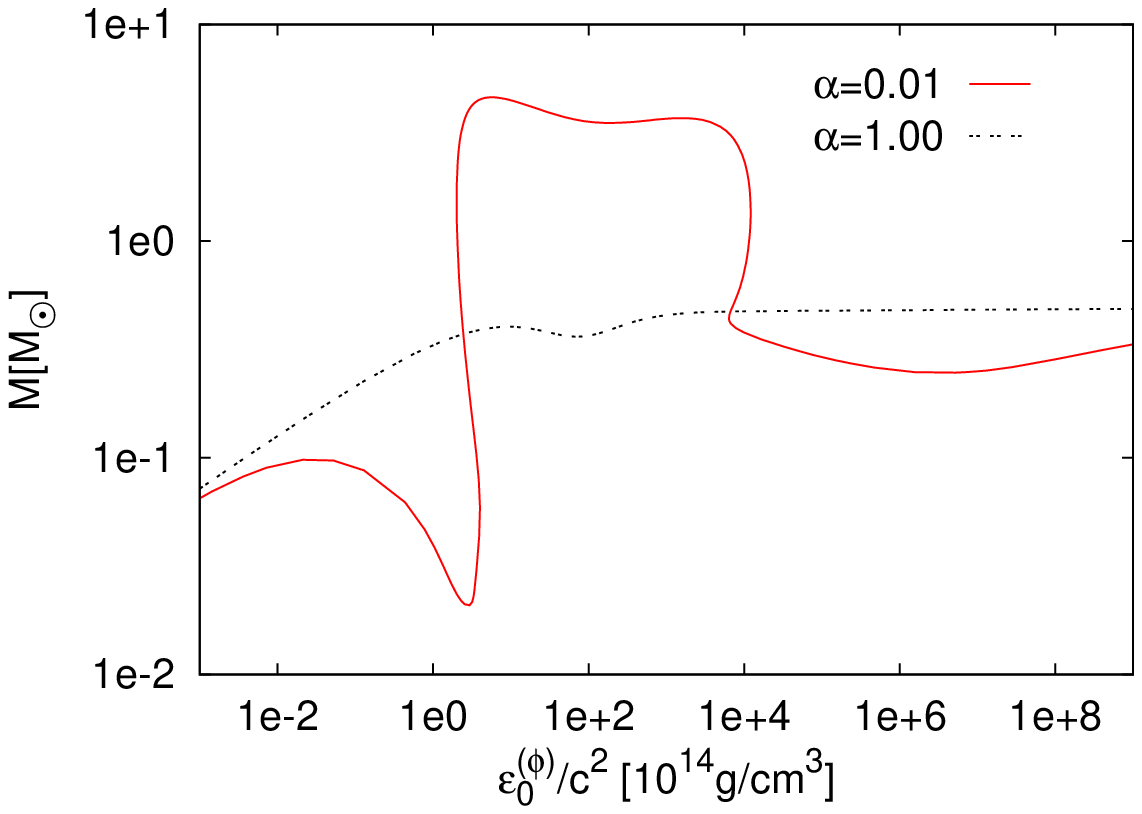}
\label{fig7b}
}
}
\end{center}
\vspace{-0.7cm}
\caption{
The mass $M$ in solar masses $M_\odot$
versus the radius $R_\star$ in km 
for families of solutions with $\alpha=0.01$ and 1,
where the insets magnify the vicinities of the limiting solution.
The corresponding Schwarzschild black hole curve is also indicated. (a)
The mass $M$ in solar masses $M_\odot$
versus the energy density of the boson field 
$\varepsilon^{(\Phi)}_{0}= - T_t^{t(\Phi)}$ at $\eta=0$,
with $\varepsilon^{(\Phi)}_{0}/c^2$ in units of $10^{14}$ g/cm$^3$
for the same set of solutions (b).
\label{fig7}
}
\end{figure}

Let us finally address some astrophysical properties of these
families of boson star solutions with nontrivial topology,
postponing the discussion of their stability to the next section.
While the mass $M$ has been considered above, we would now
like to present it in units of solar masses, $M_\odot$.
At the same time, the size $R_\star$ of these objects is of considerable
interest for a comparison with known astrophysical objects.

For boson stars, the size is not uniquely defined, because
they do not possess a sharp surface, in general.
Let us therefore adopt the definition for the radius $R_\star$
\begin{equation}
R_\star= \frac{\int j^t \left| g \right|^{1/2} R(\eta) d\eta }
     {\int j^t \left| g \right|^{1/2} d\eta } \ .
\label{radius}
\end{equation}
It has been shown previously, that the radius $R_\star$ 
of (non-compact) boson stars
is rather insensitive to the various definitions employed 
(see e.g.~\cite{Kleihaus:2011sx}).

Fig.~\ref{fig7a} shows the mass $M$ in units of the solar mass 
versus their radius $R_\star$ in km 
for such two families of boson stars with nontrivial topology,
obtained for a small and a large value of $\alpha$.
We note, that the dependence $M(R_\star )$ remains very similar
to the one known for ordinary boson stars with the
same self-interaction, in their physically
relevant range.

These boson stars possess two stable regions.
This is analogous to compact stars,
which possess a low density phase, corresponding to white dwarfs,
and a high density phase, corresponding to neutron (or quark) stars.
Furthermore, depending on the parameters employed
for the boson mass and the self-interaction, these boson stars 
can reach huge masses and get very close to the black hole limit
\cite{Kleihaus:2011sx}.

Here we see that both the low density phase and the high density
phase of those boson stars are retained 
in the presence of the nontrivial topology,
while the black hole limit is closely approached, as well.
The dependence $M(R_\star )$ is different
only close to the black hole limit,
where the solutions are expected to be
unstable on generals grounds, 
since they have passed the maximum of the mass.

Fig.~\ref{fig7b} exhibits the mass $M$ versus 
the energy density of the boson field
$\varepsilon^{(\Phi)}_{0}$ at $\eta=0$.
In the high density phase the boson field energy first assumes values
on the order of the central energy density of neutron stars.
After the maximal mass has been reached, this part of the energy density 
increases further, while the radius and the mass decrease.
Deviating from ordinary boson stars and neutron stars, however,
there occurs a subsequent strong increase of the
boson field energy density, which is compensated by the 
energy density of the phantom field, thus retaining a finite
value for the mass.

Clearly, the relevant question left to be answered for potential
astrophysical applications is the question of the stability
of these solutions. Therefore we now turn to their stability analysis.

\section{Stability Analysis}

Boson stars are stable in large regions of their parameter space.
Their stability has been investigated 
from different points of view.
Lee and Pang \cite{Lee:1988av} and Jetzer \cite{Jetzer:1991jr}
performed a linear stability analysis of boson stars
with respect to small oscillations,
while Kusmartsev, Mielke and Schunck \cite{Kusmartsev:2008py,Kusmartsev:1992},
were the first to apply catastrophe theory to boson stars.

Isolated wormholes with phantom fields, on the other hand, are unstable.
Indeed, the Ellis wormhole possesses an unstable
mode \cite{Gonzalez:2008wd,Bronnikov:2011if}, 
that had been missed before, because the gauge condition
taken had been too stringent. 
Previous investigations of configurations 
with phantom field wormholes at their core 
showed, that solutions, which are stable in the absence of 
such a wormhole become unstable in its presence
\cite{Bronnikov:2011if,Dzhunushaliev:2013lna,Charalampidis:2013ixa}.
This is, for instance, the case for astrophysical objects
like neutron stars \cite{Dzhunushaliev:2013lna},
but also for microscopic objects like Skyrmions \cite{Charalampidis:2013ixa},
which both inherit the instability of the isolated wormhole.

Here we investigate whether also boson stars inherit
the instability of the Ellis wormhole,
restricting to a linear stability analysis in the spherically symmetric sector.
We proceed by making a careful choice of the gauge condition,
which ensures that we do not miss the unstable mode present in the 
Ellis wormhole.

We start from the general spherically symmetric metric
\begin{equation}
ds^2 = -h_0(t,\eta) dt^2 + h_1(t,\eta) d\eta^2 
       +h_2(t,\eta) \left(\eta^2+r_0^2\right)d\Omega_2^2 \ ,
\end{equation}
and employ for the complex boson field and the phantom field 
the time-dependent Ans\"atze
\begin{equation}
\Phi = e^{i \omega t}\left( \Phi_1(t,\eta)+i\Phi_2(t,\eta)\right)
\ {\rm resp.}\  \ \Psi = \Psi(t,\eta) \ .
\label{skyrmUt}
\end{equation}

For these Ans\"atze we derive the Einstein-matter equations
and consider the following
perturbed metric, phantom field and boson field profile functions
\begin{eqnarray}
h_0(\eta,t)     & = & A^2 (\eta)    + \delta h_0(\eta)    e^{-i\sigma t} ,\nonumber \\
h_1(\eta,t)     & = & 1             + \delta h_1(\eta)    e^{-i\sigma t} ,\nonumber \\
h_2(\eta,t)     & = & N(\eta)       + \delta h_2(\eta)    e^{-i\sigma t} ,\nonumber \\
\Phi_1(\eta,t)  & = & \phi(\eta)    + \delta \phi_1(\eta) e^{-i\sigma t}, \nonumber \\
\Phi_2(\eta,t)  & = &               \partial_t \left(  \delta \phi_2(\eta) e^{-i\sigma t}\right), \nonumber \\
\psi(\eta,t)    & = & \psi(\eta)    + \delta \psi(\eta)   e^{-i\sigma t}. 
\end{eqnarray} 
Here $A$, $N$, $\phi$ and $\psi$ denote the unperturbed metric, boson  field
and phantom field functions.
In the next step we expand the set of Einstein-matter equations 
up to first order in the small quantities 
$\delta h_0(\eta)$, $\delta h_1(\eta)$, $\delta h_2(\eta)$, 
$\delta\phi_1(\eta)$, $\delta\phi_2(\eta)$,
and $\delta \psi(\eta)$. 
This leads to a set of linear ODEs for the perturbations, which form
an eigenvalue problem with eigenvalue $\sigma^2$. 
When $\sigma^2$ is negative
the perturbations increase in time. Thus the solution is unstable.

To reduce the number of linear ODEs we may use the gauge freedom.
Let us consider the ODE for the perturbation of the phantom field 
\begin{equation}
\left[ A N (\eta^2+r_0^2)\left(\delta\psi' 
-\frac{1}{2}    {\psi}'\left(\delta h_1-2\frac{\delta h_2}{N}-\frac{\delta h_0}{A^2}\right)\right)\right]'
+\sigma^2 \frac{N}{A} (\eta^2+r_0^2)\delta\psi = 0 \ .
\label{delphieq}
\end{equation}
This can be simplified by the gauge condition 
\begin{equation}
\delta h_1-2\frac{\delta h_2}{N}-\frac{\delta h_0}{A^2} = 0 \ .
\label{gaugecon}
\end{equation}
Furthermore, when employing (\ref{gaugecon}),
it follows that the function $\delta \psi$ vanishes identically, 
if there exists an unstable mode, i.e., if $\sigma^2$ is negative.
This is seen as follows.
Taking the gauge condition Eq.~(\ref{gaugecon}) into account we multiply
Eq.~(\ref{delphieq}) by $\delta\psi$ and integrate over
the whole range $(-\infty , \infty)$.
An integration by parts then yields
\begin{equation}
\left. \left(A N (\eta^2+r_0^2) \delta\psi  \delta\psi'\right)\right|_{-\infty}^{\infty}
  = 
     \int_{-\infty}^{\infty}\left[(\eta^2+r_0^2) \left( A N {\delta\psi'}^2 
                                     -\sigma^2 \delta\psi^2 \frac{N}{A}\right) \right]d\eta\ .
\nonumber
\end{equation}
The left-hand-side of this equation must vanish 
for a normalizable $\delta\psi$. Thus the right-hand-side must also
vanish. 
However, for a negative eigenvalue $\sigma^2$ the integrand is positive
for a finite $\delta\psi$.
The integral can only vanish, if $\delta\psi$ is identically zero.

With $\delta\psi=0$ and $\delta h_0 = A^2( \delta h_1-2\delta h_2/N)$ 
the resulting equations consist of four second order ODEs for the functions 
$\delta h_1$, $\delta h_2$, $\delta \phi_1$ and $\delta \phi_2$ in addition to two constraints $C_1$, $C_2$.
These equations are given in compact form as
\begin{equation}
\left(
\begin{array}{c} \delta h_1'' \\ \delta h_2'' \\ \delta \phi_1'' \\ \delta \phi_2'' \\ C_1 \\ C_2 \end{array}
\right) 
- \mathbf{W} 
\left(
\begin{array}{c} \delta h_1 \\ \delta h_2 \\ \delta \phi_1 \\ \delta \phi_2 \end{array}
\right) 
- \mathbf{W}^{(\eta)}
\left(
\begin{array}{c} \delta h_1' \\ \delta h_2' \\ \delta \phi_1' \\ \delta \phi_2' \end{array}
\right) 
= \left(
\begin{array}{c} 0 \\ 0 \\ 0 \\ 0 \\ 0 \\ 0\end{array}
\right)  \ ,
\label{matrices}
\end{equation}
where $\mathbf{W}$ and $\mathbf{W}^{(\eta)}$ are $6 \times 4$ matrices given in the Appendix.

\begin{figure}[t!]
\begin{center}
\vspace{0.5cm}
\mbox{\hspace{-0.5cm}
\subfigure[][]{\hspace{-1.0cm}
\includegraphics[height=.25\textheight, angle =0]{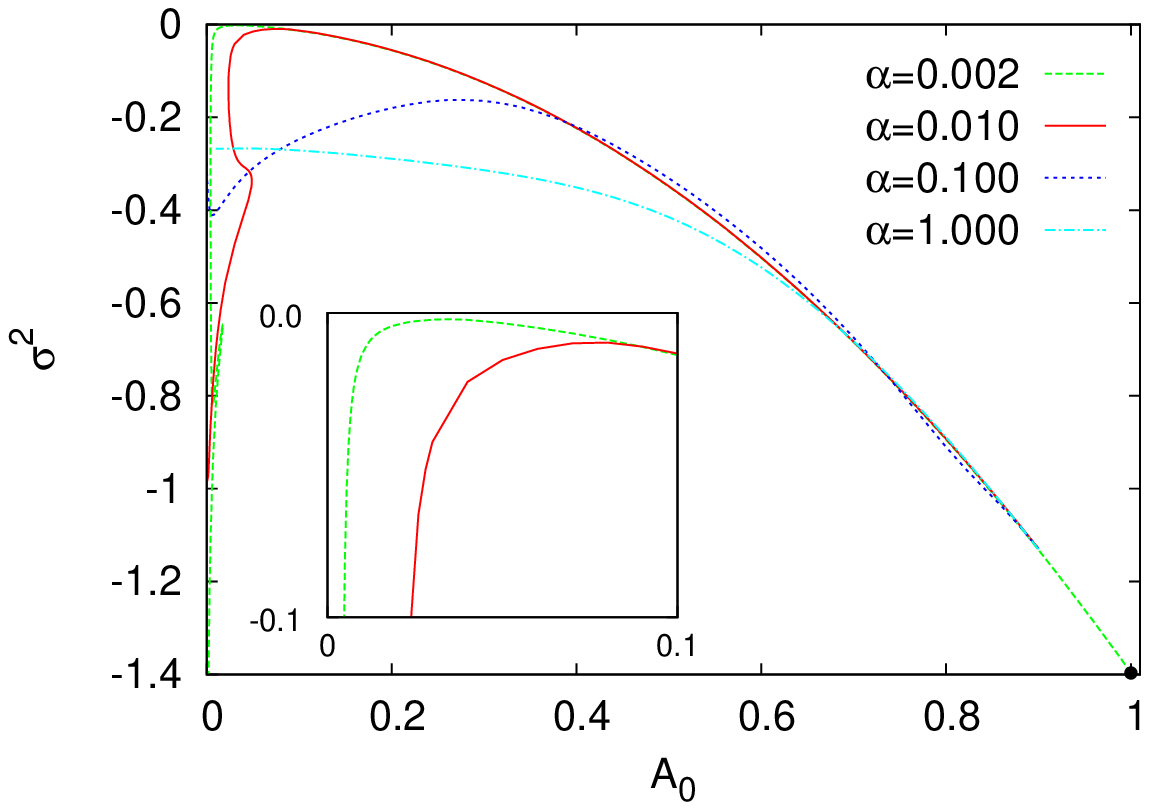}
\label{fig8a}
}
\subfigure[][]{\hspace{-0.5cm}
\includegraphics[height=.25\textheight, angle =0]{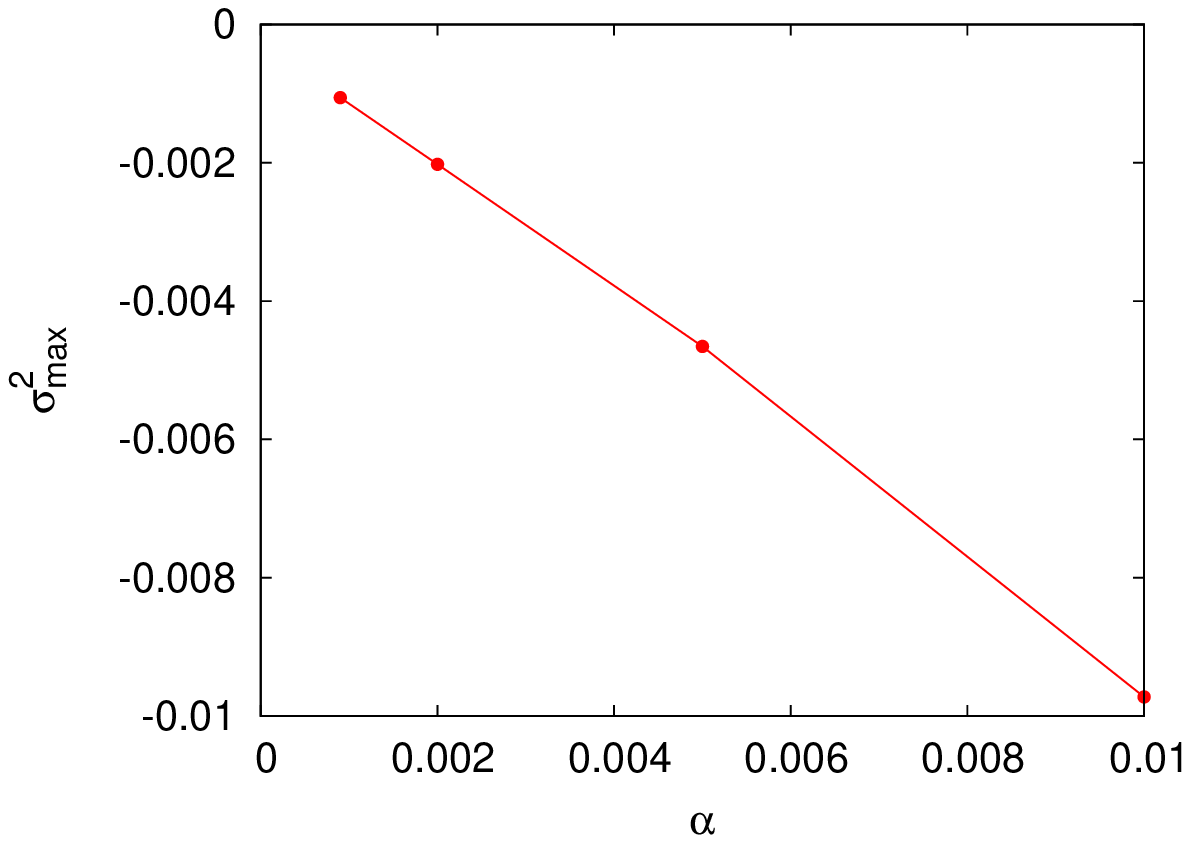}
\label{fig8b}
}
}
\end{center}
\vspace{-0.7cm}
\caption{
Instability: 
The eigenvalue $\sigma^2$ versus the value of the metric function
$A_0=A(0)$ at the throat (or equator)
for several values of $\alpha$,
where the inset magnifies the region of the eigenvalues close to zero.
The fat dot denotes the eigenvalue of the Ellis wormhole (a). 
The maximal value of the eigenvalue $\sigma^2_{\rm max}$ versus
the coupling constant $\alpha$ for $\alpha \to 0$ (b).
\label{fig8}
}
\end{figure}

Here we note that in principle the constraints
can be used to reduce the number of ODEs from four to three. However, 
this would introduce factors $1/N'$ in the ODEs, 
which diverge when $R$ becomes extremal. 
Therefore, we prefer to solve the system of the four ODEs numerically,
and to verify
that the constraints are satisfied.

The boundary conditions follow from the requirement that the 
perturbations have to vanish in the asymptotic regions,
\begin{equation}
 \delta h_1(\pm\infty)= \delta h_2 (\pm\infty) =\delta \phi_1(\pm\infty)=\delta \phi_2(\pm\infty)=0 \ .
\label{bcpert}
\end{equation}
In addition, 
we impose the condition $\delta h_2(0)=1$,
to ensure that the perturbations are normalizable.
The eigenvalue $\sigma^2$ is adjusted such that the perturbations 
satisfy the asymptotic boundary 
conditions \footnote{Technically we define an auxiliary function  
$q =\sigma^2$ and add the 
ODE: $q'=0$ to the systems of ODEs, 
without imposing a boundary condition for $q$.
Then, the number of boundary conditions of the equations in (\ref{bcpert}) 
matches the total order of the system of ODEs. 
The value of $q$ is  computed together with the solutions.}.

To make sure that we perform only 
variations with fixed particle number $Q$,
we consider its variation $ \delta Q$
\begin{equation}
\delta Q = 8\pi \int{\frac{\phi (\eta^2+r_0^2)}{A}
\left[2\omega \phi\delta h_2 
+N\left(2\delta \phi_1 -\sigma^2 \delta \phi_2\right)\right]}d\eta \ .
\end{equation}
To see that $ \delta Q$ vanishes, 
we note that the integrand can be written as a derivative,
\begin{equation}
\frac{\phi (\eta^2+r_0^2)}{A} \left[
2\omega \phi\delta h_2 
+N\left(2\delta \phi_1 -\sigma^2 \delta \phi_2\right)
\right]
= 2 \left[A N (\eta^2+r_0^2)
    \left(\delta \phi_2' \phi - \delta \phi_2 \phi'\right)\right]' \ ,
\end{equation}
provided $\delta h_1$, $\delta h_2$,  $\delta \phi_1$ and $\delta \phi_2$
are  solutions of the eigenvalue problem.
Consequently, $\delta Q = 0$ as a result of the boundary conditions
$\delta\phi_2'(0)=0$, $\phi'(0)=0$, $\delta\phi_2(\pm\infty)=0$ 
and $\phi(\pm\infty)=0$.

We demonstrate the numerical results in Fig.~\ref{fig8}. 
We show the eigenvalue $\sigma^2$ versus the value of 
the metric function $A_0$ at the throat (or equator)
in Fig.~\ref{fig8a} for several families of solutions,
covering a large range of the coupling constant $\alpha$.
The eigenvalue always starts at the value of the 
Ellis wormhole \cite{Gonzalez:2008wd}, indicated by
the fat black dot at $A_0=1$.
As $A_0$ decreases, the eigenvalue $\sigma^2$ increases,
but only up to a maximal value $\sigma^2_{\rm max}$ 
at some $\alpha$-dependent value of $A_0$.
Thus $\sigma^2$ remains negative for these families
of solutions.

Since the maximal value $\sigma^2_{\rm max}$ increases
with decreasing $\alpha$, one might hope that at some
point the instability might be lost, and stable
boson stars with nontrivial topology might arise.
However, as seen in Fig.~\ref{fig8b},
where we show $\sigma^2_{\rm max}$ versus $\alpha$ for $\alpha \to 0$,
this is not the case.
The eigenvalue can get extremely close to zero,
but it always remains negative.
Thus we conclude that all 
boson stars harbouring wormholes at their core obtained are unstable,
though the instability can become very very weak.
  
\section{Conclusions}

Coupling Einstein gravity to a complex self-interacting
boson field as well as a phantom field,
has led us to a new type of configurations,
namely boson stars harbouring a wormhole at their core,
connecting two asymptotically flat universes.
While these solutions are spherically symmetric,
they are only stationary and not static,
because the boson field carries a harmonic time-dependence.

The nontrivial topology is rendered possible by the
presence of a phantom field, which provides the necessary
violation of the energy conditions in General Relativity.
It represents the main ingredient for the Ellis wormhole
or more general wormholes in Einstein gravity.

In a first step we have considered the probe limit of the solutions,
describing boson stars in the background of an Ellis wormhole,
where only the ODE for the complex scalar field needs to be solved.
The resulting families of configurations are labelled by
the throat size $r_0$ of the background solution 
and by a parameter of the boson field, such as the frequency
$\omega$ or the value of the boson field function 
at the throat (or equator), $\phi_0$.

Interestingly, the boson frequency $\omega$ is confined
to the same interval $(\omega_{\rm min},\omega_{\rm max})$
as for $Q$-ball solutions, which are approached in the 
limit of vanishing throat size, when the background
becomes trivial and the Minkowski spacetime is recovered.
As compared to the topologically trivial $Q$-ball solutions,
the new feature appearing in the probe limit of this new type
of solutions is their nonuniqueness.
This means that above a certain value of the throat size
several distinct solutions can be found
for the same value of the frequency (within a small range).

From a theoretical point of view it is also interesting to
see how the solutions of the full gravitating system
evolve from the probe limit.
When the backreaction of the boson field on the
metric is taken into account, 
further interesting phenomena arise.
First of all, the characteristic spirals of the ordinary boson stars,
describing the dependence of their mass and particle number 
on the frequency, unwind and disappear. 
Instead all families of boson stars with nontrivial topology
extend down to zero frequency, where they
reach a singular configuration.

Second, the backreaction of the boson field effects also
strongly the geometry at the core of the configurations.
In the probe limit the solutions always feature a single throat,
but at a critical value of the coupling constant $\alpha_{\rm cr}$
the wormhole throat becomes degenerate.
Beyond $\alpha_{\rm cr}$ the wormhole then 
exhibits an equator at the core surrounded by two throats.
Thus double throat configurations emerge when the coupling
is sufficiently strong.

Further, we conclude, that such mixed configurations
of boson stars with wormholes at their core might be
of astrophysical interest. 
Depending on the choice of the potential for the complex scalar field
one can obtain such solutions with widely different masses and sizes, 
varying over many orders of magnitude.
In particular, one can also find solutions that may
mimick compact astrophysical objects like neutron stars
or black holes.
These new solutions can then serve in studies of gravitational lensing
\cite{Abe:2010ap,Toki:2011zu,Takahashi:2013jqa},
their light curves may be calculated \cite{Dzhunushaliev:2014mza}, 
the orbits of small objects in their vicinity may be determined
\cite{Eilers:2013lla}, etc.,
eventually resulting in numerous predictions for astrophysical observations.

Finally, we note that
the bosonic fields forming the boson stars
cannot fully stabilize the topologically nontrivial space-times.
The instability of the Ellis solution is inherited by the 
new configurations, composed of boson stars with wormholes at their core.
The inheritance of this instability is seen here for the first time
for stationary configurations, since the boson field
carries an explicit time-dependence.
However, depending on the parameters of the boson star potential,
the eigenvalue can get very close to zero, making the
instability very weak and allowing for long-lived configurations.

To achieve full stability two routes are suggested.
The first would be to remove the phantom field
and instead modify gravity. 
When, for instance, higher derivative or higher curvature terms
are considered,
the null energy condition can be violated
without exotic fields 
\cite{Hochberg:1990is,Fukutaka:1989zb,Ghoroku:1992tz,Furey:2004rq,Bronnikov:2009az,Kanti:2011jz,Kanti:2011yv}.
Moreover, the resulting wormholes can be stable.

The second possibility could be to include rotation.
It has been observed that the unstable mode disappears
for rotating Ellis wormholes in five dimensions,
when the rotation is sufficiently fast \cite{Dzhunushaliev:2013jja}.
Since rotating Ellis wormholes have recently also been obtained
in four spacetime dimensions \cite{Kleihaus:2014dla},
the static wormhole at the core of the configurations
could then be replaced by a rotating wormhole.
From an astrophysical point of view, 
rotating objects appear to be most relevant, anyway.

\section*{Acknowledgement}

We gratefully acknowledge support by the German Research Foundation
within the framework of the DFG Research Training Group 1620
{\it Models of gravity}
as well as support by the Volkswagen Stiftung.
VD and VF
gratefully acknowledge a grant in fundamental research in natural sciences
by the Ministry of Education and Science of Kazakhstan
for the support of this research.

\section{Appendix}

Here we present the $6 \times 4$ matrices 
$\mathbf{W}$ and $\mathbf{W}^{(\eta)}$ occurring in Eq.~(\ref{matrices})
of the stability analysis:
\begin{eqnarray}
\mathbf{W}_{11} & = & - 4 \alpha \lambda  (b + \phi^4 - c \phi^2) \phi^2  -\frac{\sigma^2}{A^2}
\nonumber\\
\mathbf{W}_{12} & = & \frac{2 (([\eta^2+r_0^2] N' + 2 N \eta) A - 2 A' [\eta^2+r_0^2] N) N'}{A [\eta^2+r_0^2] N^3}
\nonumber\\
\mathbf{W}_{13} & = &  - 8 \alpha \lambda (b + 3 \phi^4 - 2 c \phi^2) \phi 
\nonumber\\
\mathbf{W}_{21} & = & - 2\frac{ 2 \alpha \lambda  N(b + \phi^4 - c \phi^2) [\eta^2+r_0^2] \phi^2  - 1}{\eta^2+r_0^2}
\nonumber\\
\mathbf{W}_{22} & = & - ((4 \alpha \lambda N^2(b + \phi^4 - c \phi^2) [\eta^2+r_0^2] \phi^2 + 2 (N + N' \eta) N 
                    + [\eta^2+r_0^2] N'^2) A^2 
\nonumber\\
& &                  + 2 A A' \eta N^2)/(A^2 [\eta^2+r_0^2] N^2)
                        -\frac{\sigma^2}{A^2}
\nonumber\\
\mathbf{W}_{23} & = &  - 8 \alpha \lambda N (b + 3 \phi^4 - 2 c \phi^2)  \phi 
\nonumber\\
\mathbf{W}_{31} & = &  \lambda (b + 3 \phi^4 - 2 c \phi^2) \phi
\nonumber\\
\mathbf{W}_{32} & = & - 2 \frac{\phi \omega^2}{A^2 N}
\nonumber\\
\mathbf{W}_{33} & = & \frac{\lambda  A^2 (b + 15 \phi^4 - 6 c \phi^2)  - \omega^2}{A^2}-\frac{\sigma^2}{A^2}
\nonumber\\
\mathbf{W}_{34} & = & 2 \frac{\omega \sigma^2}{A^2}
\nonumber\\
\mathbf{W}_{42} & = & 2 \frac{\phi \omega}{A^2 N}
\nonumber\\
\mathbf{W}_{43} & = & 2 \frac{\omega}{A^2}
\nonumber\\
\mathbf{W}_{44} & = & \frac{\lambda A^2 (b + 3 \phi^4 - 2 c \phi^2) - \omega^2 }{A^2}-\frac{\sigma^2}{A^2}
\nonumber\\
\mathbf{W}_{51} & = & 4 (2\alpha \lambda N  (b + \phi^4 - c \phi^2) [\eta^2+r_0^2] \phi^2 - 1) A^4 [\eta^2+r_0^2] N
\nonumber\\
\mathbf{W}_{52} & = &  - 2 \frac{A^2}{N} (2 ((4 \alpha \omega^2 \phi^2 - \sigma^2) N + A A' N') [\eta^2+r_0^2] N - (2 (N' \eta + 1) N 
+ [\eta^2+r_0^2] N'^2) A^2) [\eta^2+r_0^2]
\nonumber\\
\mathbf{W}_{53} & = & 16 \alpha  A^2 N^2 (\lambda  A^2 (b + 3 \phi^4 - 2 c \phi^2) - \omega^2) [\eta^2+r_0^2]^2 \phi
\nonumber\\
\mathbf{W}_{54} & = & 16 \alpha \omega \sigma^2 A^2 N^2 [\eta^2+r_0^2]^2 \phi
\nonumber\\
\mathbf{W}_{61} & = &  -  A^2 N([\eta^2+r_0^2] N' + 2 N \eta)
\nonumber\\
\mathbf{W}_{62} & = &  - ( A ([\eta^2+r_0^2] N' - 2 N \eta)+ 2 A' [\eta^2+r_0^2] N) A
\nonumber\\
\mathbf{W}_{63} & = & 8 \alpha A^2 N^2 [\eta^2+r_0^2] \phi' 
\nonumber\\
\mathbf{W}_{64} & = &  - 8 \alpha\omega A^2 N^2 [\eta^2+r_0^2] \phi'
\nonumber
\end{eqnarray}
\begin{eqnarray}
\mathbf{W}^{(\eta)}_{11} & = & \frac{  A([\eta^2+r_0^2] N' + 2 N \eta) - A' [\eta^2+r_0^2] N}{A N [\eta^2+r_0^2]}
\nonumber\\
\mathbf{W}^{(\eta)}_{12} & = & - 2 \frac{  A([\eta^2+r_0^2] N' + 2 N \eta) - 2 A' [\eta^2+r_0^2] N}{A N^2 [\eta^2+r_0^2]}
\nonumber\\
\mathbf{W}^{(\eta)}_{22} & = & \frac{ A([\eta^2+r_0^2] N' - 2 N \eta) - A' [\eta^2+r_0^2] N}{A N [\eta^2+r_0^2]}
\nonumber\\
\mathbf{W}^{(\eta)}_{33} & = &  - \frac{ A ([\eta^2+r_0^2] N' + 2 N \eta) + A' [\eta^2+r_0^2] N}{A N [\eta^2+r_0^2]}
\nonumber\\
\mathbf{W}^{(\eta)}_{44} & = &  - \frac{ A ([\eta^2+r_0^2] N' + 2 N \eta) + A' [\eta^2+r_0^2] N}{A N [\eta^2+r_0^2]}
\nonumber\\
\mathbf{W}^{(\eta)}_{51} & = & 2  A^4 N ([\eta^2+r_0^2] N' + 2 N \eta) [\eta^2+r_0^2]
\nonumber\\
\mathbf{W}^{(\eta)}_{52} & = &  - 2 ( A ([\eta^2+r_0^2] N' + 2 N \eta) - 2 A' [\eta^2+r_0^2] N) A^3 [\eta^2+r_0^2]
\nonumber\\
\mathbf{W}^{(\eta)}_{53} & = &  - 16 A^4 N^2 \alpha [\eta^2+r_0^2]^2 \phi'
\nonumber\\
\mathbf{W}^{(\eta)}_{62} & = & 2 A^2 N [\eta^2+r_0^2]
\nonumber\\
\mathbf{W}^{(\eta)}_{64} & = & 8 \omega A^2 N^2 \alpha [\eta^2+r_0^2] \phi 
\nonumber
\end{eqnarray}

\end{document}